\documentclass[fleqn,usenatbib]{mnras}

\usepackage{newtxtext,newtxmath}

\usepackage[T1]{fontenc}
\usepackage{ae,aecompl}
\usepackage{siunitx}

\usepackage{graphicx}	
\usepackage{amsmath}	
\usepackage{threeparttable}
\usepackage{pdflscape}

\newcommand{\hi}{\ion{H}{i}}
\newcommand{\gs}{${\rm M_{\ion{H}{i}}/{M_*}}$}
\newcommand{\oiha}{[\ion{O}{i}]/H$\alpha$}
\newcommand{\siiha}{[\ion{S}{ii}]/H$\alpha$}
\newcommand{\niiiha}{[\ion{N}{ii}]/H$\alpha$}
\newcommand{\oiiihb}{[\ion{O}{iii}]/H$\beta$}

\title[HI-MaNGA]{HI-MaNGA: Tracing the physics of the neutral and ionized ISM with the second data release}

\author[D. V. Stark et al.]{David V. Stark,$^{1}$\thanks{E-mail: dstark@haverford.edu}
Karen L. Masters,$^{1}$
Vladimir Avila-Reese,$^{2}$
Rogemar Riffel,$^{3,4}$
\newauthor
Rogerio Riffel,$^{3,5}$
Nicholas Fraser Boardman,$^{6}$
Zheng Zheng,$^{7}$
Anne-Marie Weijmans,$^{8}$
\newauthor
Sean Dillon,$^{9}$
Catherine Fielder,$^{10}$
Daniel Finnegan,$^{11}$
Patricia Fofie,$^{12}$
\newauthor
Julian Goddy,$^{1}$
Emily Harrington,$^{13}$
Zachary Pace,$^{14}$
Wiphu Rujopakarn,$^{15,16,17}$
\newauthor
Nattida Samanso,$^{15}$
Shoaib Shamsi,$^{1}$
Anubhav Sharma,$^{1}$
Elizabeth Warrick,$^{13}$
\newauthor
Catherine Witherspoon,$^{14}$
Nathan Wolthuis,$^{1}$
\\
$^{1}$Department of Physics and Astronomy, Haverford College, 370 Lancaster Avenue, Haverford, PA 19041, USA\\ 
$^{2}$Instituto de Astronom\'ia, Universidad Nacional Aut\'onoma de M\'exico, A.P. 70-264, 04510, Mexico, D.F., M\'exico\\ 
$^{3}$Laborat\'orio Interinstitucional de e-Astronomia, 77 Rua General Jos\'e Cristino, Rio de Janeiro, 20921-400, Brasil \\ 
$^{4}$Departamento de F\'isica, CCNE, Universidade Federal de Santa Maria, 97105-900, Santa Maria, RS, Brazil\\ 
$^{5}$Universidade Federal do Rio Grande do Sul, IF, CP 15051, Porto Alegre
91501-970, RS, Brazil \\ 
$^{6}$Department of Physics and Astronomy, University of Utah, 115 S. 1400 E., Salt Lake City, UT 84112, USA	\\ 
$^{7}$National Astronomical Observatories of China, Chinese Academy of Sciences, 20A Datun Road, Chaoyang District, Beijing 100012, China\\	
$^{8}$School of Physics and Astronomy, University of St Andrews, North Haugh, St. Andrews KY16 9SS, UK	
$^{9}$Department of Physics and Astronomy, University of Texas at San Antonio, One UTSA Circle, San Antonio, TX 78249, USA \\ 
$^{10}$PITT PACC, Department of Physics and Astronomy, University of Pittsburgh, Pittsburgh, PA 15260, USA	\\ 
$^{11}$Department of Physics, Siena College, 515 Loudon Road, Loudonville, New York 12211, USA \\ 
$^{12}$Williams College Department of Physics, 33 Lab Campus Drive, Williamstown, MA 01267, USA \\ 
$^{13}$Department of Physics, Bryn Mawr College, 101 N Merion Ave, Bryn Mawr, Pennsylvania 19010, USA \\ 
$^{14}$Department of Astronomy, University of Wisconsin-Madison, 475N. Charter St., Madison WI 53703, USA	\\ 
$^{15}$Department of Physics, Faculty of Science, Chulalongkorn University, 254 Phayathai Road,
Pathumwan, Bangkok 10330, Thailand \\ 
$^{16}$National Astronomical Research Institute of Thailand, Don Kaeo, Mae Rim, Chiang Mai 50180, Thailand \\ 
$^{17}$Kavli IPMU, Todai Institutes for Advanced Study (WPI),The University of Tokyo, Kashiwa, Japan 277- 8583 
}

\date{Accepted XXX. Received YYY; in original form ZZZ}

\pubyear{2020}

\begin{document}
\label{firstpage}
\pagerange{\pageref{firstpage}--\pageref{lastpage}}
\maketitle

\begin{abstract}
We present the second data release for the HI-MaNGA programme of {\hi} follow-up observations for the SDSS-IV MaNGA survey. This release contains measurements for 3669 unique galaxies, combining 2108 Green Bank Telescope observations with an updated crossmatch of the MaNGA sample with the ALFALFA survey. We combine these data with MaNGA spectroscopic measurements to examine relationships between \hi-to-stellar mass ratio (\gs) and average ISM/star formation properties probed by optical emission lines. {\gs} is very weakly correlated with the equivalent width of H$\alpha$, implying a loose connection between the instantaneous star formation rate and the HI reservoir, although the link between {\gs} and star formation strengthens when averaged even over only moderate timescales ($\sim$30 Myrs). Galaxies with elevated {\hi} depletion times have enhanced {\oiha} and depressed $H\alpha$ surface brightness, consistent with more {\hi} residing in a diffuse and/or shock heated phase which is less capable of condensing into molecular clouds. Of all optical lines, {\gs} correlates most strongly with oxygen equivalent width, EW(O), which is likely a result of the existing correlation between {\gs} and gas-phase metallicity. Residuals in the \gs$-$EW(O) relation are again correlated with [\ion{O}{i}]/H$\alpha$ and H$\alpha$ surface brightness, suggesting they are also driven by variations in the fraction of diffuse and/or shock-heated gas. We recover the strong anti-correlation between {\gs} and gas-phase metallicity seen in previous studies. We also find a relationship between {\gs} and [\ion{O}{i}]6302/H$\alpha$, suggesting that higher fractions of diffuse and/or shock-heated gas are more prevalent in gas-rich galaxies.
\end{abstract}

\begin{keywords}
galaxies: ISM -- radio lines: galaxies -- catalogs -- surveys 
\end{keywords}



\section{Introduction}
\label{sec:intro}

Neutral atomic hydrogen (\hi) is typically the dominant phase of cold gas found within galaxies at the present epoch and plays an important role in the evolution of galaxies.  Although stars are not expected to form out of \ion{H}{i} directly, it is still the large-scale gas reservoir out of which the dense, cold, star-forming molecular clouds condense \citep[e.g.][]{Elmegreen93}. Generally, galaxies with higher {\hi}-to-stellar mass ratios (\gs) have higher specific star formation rates, sSFR \citep{Doyle06,Wang11,Huang12}. {\gs} has a particularly strong relationship with galaxy color, which traces the relative growth rates of galaxies over $\sim$100 Myr to Gyr timescales \citep{Kannappan13}, such that galaxies with bluer colors/higher long-term growth rate tend to have large \ion{H}{i} reservoirs, while galaxies with red colors/lower long-term growth rates tend to have little or no detectable \ion{H}{i} \citep{Kannappan04,Catinella13,Kannappan13,Eckert15,Catinella18}. The meaning of the {\hi}-color relationship has been called into question by \citet{Jaskot15} however, who argue it is primarily driven by internal dust extinction, not recent star formation.  

The ability of {\hi} reservoirs to contribute to star formation is strongly linked to the local conditions within the interstellar medium (ISM), namely those which promote the formation of molecular clouds \citep[e.g.][]{Krumholz09,Krumholz09b}. The typical conditions of the ISM likely vary between low stellar mass (typically \hi-rich) galaxies and high stellar mass (typically \hi-poor) galaxies, which can in turn impact the ability of their {\hi} reservoirs to contribute to star formation. For instance, \citet{Dalcanton04} find that the ISM is more confined to dense thin disks in high-mass galaxies, while the ISM is more diffuse with a higher vertical scale length, in lower mass galaxies. This result further implies lower turbulent velocities and more disk instabilities in higher mass galaxies. Consistently, several studies argue low-mass galaxies are generally stable against fragmentation \citep{Meurer96,vanZee97,Hunter98}. Furthermore, spectroscopic observations have also shown that gas-phase metallicity and {\hi} content are inversely related \citep{Lequeux79, Skillman96, Peeples08, Robertson12, Moran12, Hughes13, Bothwell13, Brown18,Zu20}. The combined dynamical stability and lower metal content of low mass gas rich galaxies is expected to make it harder for them to meet the conditions necessary for molecular cloud formation and subsequent star formation.  

Optical spectroscopy provides a powerful means of probing how the conditions of the ISM vary between gas-rich and gas-poor galaxies, not only by revealing metal abundances, but also ionization fractions, properties of ionizing sources, gas densities, and gas temperatures. With the exception of metallicity, many of these properties have not been explored in great detail, and prior studies which relate gas content to optical spectroscopic properties have been limited in a number of ways. Traditionally, spatially resolved spectroscopy measured over large fractions of galaxy disks has been difficult to obtain, and such data have been limited to small samples. Sample sizes can be dramatically increased by using large fibre-fed optical spectroscopic surveys \citep[e.g., the Sloan Digital Sky Survey, SDSS,][]{York00}, but these observations only capture spectra in the nuclei of nearby galaxies. The current generation of large integral field unit (IFU) surveys like the Mapping Nearby Galaxies at Apache Point Observatory \citep[MaNGA;][]{Bundy15}, Sydney-AAO Multi-object Integral field spectrograph \citep[SAMI;][]{Croom12}, and Calar Alto Legacy Integral Field Area \citep[CALIFA;][]{Sanchez12} surveys address these limitations by providing spatially resolved spectroscopy out to large radii for thousands of galaxies.

Complementing the optical IFU observations of $\sim$10,000 galaxies in the SDSS-IV MaNGA survey, HI-MaNGA \citep{Masters19} is an \ion{H}{i} follow-up program collecting single-dish 21cm data for all MaNGA galaxies lacking coverage from other {\hi} surveys.  In this paper, we present the second data release for HI-MaNGA which contains 2108 Robert C. Byrd Green Bank Telescope (GBT) observations (a factor of 6.5 times larger than the previous data release) and an updated crossmatch against the Arecibo Legacy Fast ALFA (ALFALFA) survey \citep{Haynes18}, yielding a sample of 3669 unique galaxies with global {\hi} and optical IFU spectroscopic data. We use this powerful data set to investigate how the average ISM properties vary as a function of gas-richness, providing additional important clues as to what affects the ability of different types of galaxies to convert their ISM into stars. 

In Section~\ref{sec:data}, we present our {\hi} observations and updated catalog, a discussion of how {\hi} upper limits are incorporated into our analysis, and a description of our analysis sample and derived data products. In Section~\ref{sec:results}, we present scaling relations between {\gs} and ISM properties derived from MaNGA optical spectroscopy. In Section~\ref{sec:discussion}, we interpret our findings and discuss potential systematic errors.  Our conclusions are summarized in Section~\ref{sec:conclusions}.

Throughout this work, we assume a cosmology with $H_0=70\,{\rm km\,s^{-1}\,Mpc^{-1}}$, $\Omega_m=0.3$, and $\Omega_{\Lambda}=0.7$.

\section{Data and Methods}
\label{sec:data}
\subsection{New HI-MaNGA Data}
\label{sec:new_hi_data}

HI-MaNGA \citep{Masters19} is an {\hi} follow-up program for the SDSS-IV Mapping Nearby Galaxies at Apache Point Observatory (MaNGA) survey \citep{Gunn06,Smee13,Bundy15,Blanton17,Yan16a}. In this paper, we release GBT observations of 2108 galaxies that were taken over 1272 hours between 2016 and 2018 (proposal codes AGBT16A-095 and AGBT17A-012). Details about target selection, observing strategy, and data reduction can be found in the data release 1 (DR1) paper \citep{Masters19}. To briefly summarize, all MaNGA galaxies lacking {\hi} data are targeted, with no pre-selection based on other properties such as stellar mass, color, or morphology, although there is a $z<0.05$ restriction which introduces a bias against the highest stellar mass galaxies due to the stellar mass-redshift relation built into the MaNGA sample definition \citep{Wake17}. All targets are observed using standard position-switching to the same depth ($\sim$1.5 mJy, $\sim$15 minutes on-source). All scans are averaged, any strong radio frequency interference (RFI) is removed, and the spectra are boxcar and hanning smoothed to a final velocity resolution of ${\sim}10\,{\rm km\,s^{-1}}$. A polynomial is fit to remove any leftover baseline variations, after which source parameters are measured (see Section~\ref{sec:catalog}).

There are few modifications to the derived data products in this release compared to DR1. First, we have incorporated cosmological corrections (factors of $1+z$) to relevant quantities. Second, all fluxes have been multiplied by a factor of 1.2 to correct for an underestimation of the default flux calibration scale built into \texttt{GBTIDL}\footnote{\url{http://gbtidl.nrao.edu/}}, the data reduction software used to process our data (see \citealt{Goddy20} for further details). Third, linewidths have been corrected for instrumental broadening. Finally, the catalog contains additional information to aid users, including additional general information for galaxies, an assessment of source confusion, and flags to highlight potentially unreliable observations. Further details are given in Section~\ref{sec:catalog}.

\subsection{ALFALFA crossmatch}
\label{sec:alfalfa_crossmatch}
A significant fraction of MaNGA ($\sim$30\%) overlaps with the ALFALFA survey \citep{Haynes18}. In the second HI-MaNGA data release, we include a crossmatch between ALFALFA and the MaNGA Product Launch 8 (MPL-8) sample. To match the ALFALFA catalog (a.100) with MaNGA galaxies, we first crossmatch with the NASA Sloan Atlas (NSA\footnote{\url{http://www.nsatlas.org/}}, which includes all MaNGA galaxies).  All optical counterparts within one half-power beam width (HPBW) to each {\hi} source are matched to it, although cases with multiple optical matches, including any out to 1.5 times the HPBW, are flagged as potentially confused. Further details on the matching process, including how matching is done in velocity space, are given in Section~\ref{sec:confusion}. For any MaNGA galaxies which fall inside the ALFALFA footprint but do not have matches within the a.100 catalog, we extract its spectrum from the ALFALFA data cubes using a $4\arcmin\times4\arcmin$ window at each galaxy's location. A first order baseline is subtracted, and the rms noise is measured enabling an estimate of an {\hi} flux upper limit. Unlike the GBT observations, we do not explicitly limit the ALFALFA crossmatch to $z<0.05$, although the outer redshift limit of ALFALFA does not extent significantly beyond this range.

ALFALFA is an untargeted {\hi} survey, and as such requires a relatively high signal-to-noise ratio ($S/N$) for detections in order to consider them reliable enough to incorporate into the final a.100 catalog. In contrast, a pointed survey (like our GBT observations) can accept lower $S/N$ on detections due to the prior knowledge of where the galaxies are located both on the sky and in redshift space. The a.100 catalog contains code 1 sources with $S/N>6.5$, and code 2 sources with lower $S/N$ if they coincide with an existing optical counterpart with known redshift. Although there is no minimum $S/N$ required for the code 2 sources, their frequency declines rapidly below $S/N\sim4.5$. This $S/N$ corresponds to an integrated flux approximately matching the ``detection limit" of the survey for a linewidth of 200 km s$^{-1}$ \citep{Haynes11}. While non-detections in our GBT observations are used to derive $3\sigma$ upper limits, using $3\sigma$ upper limits for objects lacking matches in the a.100 catalog would be inappropriate because there is the possibility that they are weak sources with $3 < S/N < 4.5$ that were missed. Therefore, all ALFALFA non-detections are treated as $4.5\sigma$ upper limits in order to more accurately match the completeness level of the a.100 catalog. 

\subsection{Source Confusion}
\label{sec:confusion}

The GBT and Arecibo beams are 9$\arcmin$ and 3.5$\arcmin$, respectively, at 21cm. Thus, there is a strong potential for {\hi} source confusion caused by multiple galaxies within the beam.  We now flag potential cases of confusion for all {\hi} detections by identifying all known galaxies from the NSA within 1.5 times the HPBW from the beam center and at similar velocity as our primary target. Specifically, assuming the primary galaxy (that being targeted) has a linewidth $W_{50}$ and central velocity $V_{HI}$ equivalent to what was measured, and the secondary galaxies (those in close proximity to the primary) have linewidths of $W_{50}=200\,{\rm km\,s^{-1}}$ and central velocities equivalent to their known optical redshifts, we look for overlap between their {\hi} profiles in velocity space. Any cases with velocity overlap are flagged as potentially confused (see the \texttt{confused} column in the final catalog (Section~\ref{sec:catalog}). For this analysis, we assess velocity overlap using $W_{50}+20\,{\rm km\,s^{-1}}$ to account for the fact that profiles typically extend slightly beyond $W_{50}$. This adjustment is more akin to using a $W_{20}$ measurement \citep{Kannappan02}, but we use the approximation instead of actual measurements of $W_{20}$ because the measurements may be unreliable for a significant fraction of our data set with low $S/N$. We also enforce a minimum $W_{50}$ of 20 km s$^{-1}$ to account for turbulent motions. 

The above analysis attempts to identify all cases of confusion, but likely over-estimates the true rate of confusion because some secondary galaxies may be inherently gas-poor relative to the primary target and/or far enough away from the beam center that they do not contribute significantly to the total measured {\hi} flux. To refine our estimate of source confusion, we perform an additional analysis taking into account each galaxy's likely {\hi} mass and position relative to the beam center. For each potentially confused observation (identified as described above), we estimate the likelihood distributions of $M_{HI}$ for the primary and secondary galaxies using their color and surface brightness as described in Section~\ref{sec:photogs} and Appendix~\ref{app:photogs}. We estimate the relative flux likelihood distributions of each galaxy by scaling the {\hi} mass likelihood distributions based on the relative beam power at their angular distance from the beam center, where this scale factor is estimating assuming a Gaussian beam with full-width-half-max (FWHM) appropriate for Arecibo of GBT observations. We then calculate the likelihood distribution of the ratio, $R$, of the flux, $F_{\rm HI}$, from from all secondary galaxies to that of all galaxies:
\begin{equation}
R = \frac{\sum F_{\rm HI,secondary}}{F_{\rm HI,primary} + \sum F_{\rm HI,secondary}}
\end{equation}
We then estimate $P_{R>0.2}$, the probability that $R$ is greater than 0.2, i.e., the probability that the companions are contributing at least 20\% of the total measured flux (a value of 20\% is chosen as it is comparable to typical flux calibration uncertainties). $P_{R>0.2}$ serves as means of incorporating galaxies back into an analysis which likely have reliable flux measurements even in the presence of nearby companions. For example, only considering galaxies with $P_{R>0.2} > 0.1$ as confused lowers the total confusion rates of GBT and ALFALFA data by 12\% and 22\%, respectively. We incorporate $P_{R>0.2}$ into our catalog (Section~\ref{sec:catalog}) so that any user can make their own judgment as to what value is appropriate for their analysis. For our analysis, we adopt the threshold of $P_{R>0.2}>0.1$, which removes 526 galaxies from the full catalog (105 from ALFALFA and 421 from GBT).

\subsection{HI-MaNGA DR2 Catalog and Products}
\label{sec:catalog}
The HI-MaNGA second data release (DR2) catalog can be found at \url{https://greenbankobservatory.org/science/gbt-surveys/hi-manga/}. This release includes the galaxies from HI-MaNGA DR1 \citep{Masters19}, and all data products here supersede those from the earlier catalogue. The data will also be released as part of a Value Added Catalogue with SDSS-IV DR17, which will enable easy access alongside all MaNGA data through the \texttt{Marvin} python package \citep{Cherinka19}. 

The DR2 catalog contains {\hi} information for MaNGA galaxies in the MaNGA Product Launch (MPL) 8 sample, although does not provide {\hi} data for {\it all} 6583 galaxies in MPL-8. The full catalog contains 3818 galaxies, including 1809 detections and 2009 non-detections. There are 2108 GBT observations, with 1191 detections and 917 non-detections (this count includes the 331 galaxies from DR1). The remaining 1710 galaxies are those with ALFALFA data, of which 618 are detections (with counterparts found in the a.100 catalog; \citealt{Haynes18}) and 1092 are non-detections. For the GBT data, a ``detection" is judged by eye, with no explicit minimum $S/N$, although each reduced spectrum and derived quantities are inspected by a second person for quality control. The lowest integrated $S/N$ (defined as the total flux divided by the total flux uncertainty) among the GBT data is 1.4, although such low values are exceptionally rare; 99\% of GBT detections have integrated $S/N>3$. As discussed in Section~\ref{sec:alfalfa_crossmatch}, ALFALFA does not enforce a minimum $S/N$ (defined with Eq. 4 in \citealt{Haynes18}), but detections are very rare below $S/N\sim4.5$. There are 149 galaxies with both GBT and ALFALFA data (these GBT data were mostly taken prior to the final release of the ALFALFA catalog when the final footprint was unknown), and we include both observations for these galaxies in this catalogue. See Section~\ref{sec:analysis_sample} for a description of the criteria used in this work to select which observation is incorporated into our final sample.

A complete description of the quantities in the final catalog is as follows:
\begin{enumerate}
    \item[\bf MaNGA-ID, Plate-IFU:] Internal MaNGA-ID and Plate-IFU designations. \newline
    \item[$\boldsymbol{\alpha}$,$\boldsymbol{\delta}$:] Galaxy Right Ascension and Declination, J2000 (taken from MaNGA DRPALL catalog\footnote{\url{https://www.sdss.org/dr16/manga/manga-data/catalogs/}}).\newline
    \item[$\boldsymbol{\log{M_*}}$:] Base-10 log of the stellar mass. Taken from the DRPALL catalog, which is turn is taken from the NSA catalog.\newline
    \item[$\boldsymbol{\sin{i}}$:] An estimate of the sin of the inclination using the axial ratio, $b/a$:
    \begin{equation}
        \sin{i}=\sqrt{\frac{1-(b/a)^2}{1-q^2}}
    \end{equation}
    where $q=0.2$ is the assumed intrinsic thickness and $b/a$ is taken from the DRPALL catalog.  \newline 
    \item[$\boldsymbol{V_{\rm opt}}$:] Systemic velocity derived from the optical spectrum of this target (taken from MaNGA DRPALL catalog).\newline
    \item[\bf Session:] Session IDs during which observations were conducted. For GBT observations, the format is \texttt{[program ID]\_[session\_number]} (e.g., \texttt{AGBT17A\_012\_01}). For all data from the ALFALFA survey, the session is listed as \texttt{ALFALFA}.\newline
    \item[$\boldsymbol{t_{int}}$:] Total on-source integration time in seconds. \newline
    \item[$\boldsymbol{rms}:$] rms noise in mJy measured in a signal free part of the {\hi} spectrum. \newline
    \item[$\boldsymbol{\log{M_{\rm HI,lim,200kms}}}$] For non-detections, the base-10 log of the {\hi} mass upper limit (in $M_{\odot}$). For GBT data, this is calculated as
    \begin{equation}
        \frac{M_{HI}}{M_{\odot}} = 3\times\frac{2.36\times10^5}{(1+z)^{2}}\left(\frac{D}{{\rm Mpc}}\right)^2\left(\frac{rms}{Jy}\right)\sqrt{\left(\frac{W}{\rm km\,s^{-1}}\right)\left(\frac{dV}{\rm km\,s^{-1}}\right)}
    \end{equation}
    where $W$ is the assumed linewidth, $D$ is the luminosity distance to the galaxy, and $dV$ is the velocity resolution. For ALFALFA data, the upper limits are estimated by reordering Equation (4) from \citet{Haynes18} to determine the flux of a $S/N=4.5$ detection (4.5 being the minimum allowed $S/N$ for an detection with an existing optical redshift in the a.100 catalog), then converting to HI mass:
    \begin{equation}
        \frac{M_{HI}}{M_{\odot}} = 4.5 \times \frac{2.36\times10^5}{(1+z)^{2}}\left(\frac{D}{{\rm Mpc}}\right)^2 \left(\frac{W}{\rm km\,s^{-1}}\right)\left(\frac{rms}{\rm Jy}\right)w_{smo}^{-1/2}
    \end{equation}
    where $w_{smo}=W/20$. We assume $W=200\,{\rm km\,s^{-1}}$ for all upper limits.\newline
    \item[$\boldsymbol{f_{\rm peak}}$:]  The peak flux density of the detected {\hi} emission line.\newline
    \item[$\boldsymbol{S/N}$:] The peak signal-to-noise ratio defined as $f_{\rm peak}/rms$.\newline
    \item[$\boldsymbol{F_{\rm HI}}$] For detections, the integrated flux of the {\hi} line in Jy km s$^{-1}$.\newline
    \item[$\boldsymbol{\log{M_{HI}}}$:] For detections, base-10 log of the {\hi} mass (in solar masses) defined as
    \begin{equation}
        \frac{M_{HI}}{M_{\odot}} = \frac{2.36\times10^5}{(1+z)^{2}}\left(\frac{D}{\rm Mpc}\right)^2\left(\frac{F_{HI}}{\rm Jy\,km\,s^{-1}}\right), 
    \end{equation}
    \newline
    \item[$\boldsymbol{W_{M50}}$:] Width of the {\hi} line measured at 50\% of the mean flux density of the two peaks (or the overall peak flux density in the case of a single-peaked profile). The locations of the 50\% peak flux density levels on either side of the {\hi} profile are estimated using linear interpolation. The width is corrected for instrumental broadening and cosmological stretch. \newline
    
    \item[$\boldsymbol{W_{P50}}$:] Width of the {\hi} line measured at 50\% of the overall peak flux density of the {\hi} line. The locations of the 50\% peak flux density levels on either side of the {\hi} profile are estimated using linear interpolation. The width is corrected for instrumental broadening and cosmological stretch.\newline
    
    \item[$\boldsymbol{W_{P20}}$:] Width of the {\hi} line measured at 20\% of the overall peak flux density of the {\hi} line. The locations of the 20\% peak flux density levels on either side of the {\hi} profile are estimated using linear interpolation. The width is corrected for instrumental broadening and cosmological stretch.\newline
    
    \item[$\boldsymbol{W_{2P50}}$:] Width of the {\hi} line measured at 50\% of the peak flux density on either side of the profile (or the overall peak flux density in the case of a single-peaked profile). The location of the 50\% peak flux density level is estimated using linear interpolation. The width is corrected for instrumental broadening and cosmological stretch.\newline
    
    \item[$\boldsymbol{W_{F50}}$:] Width of the {\hi} line measured at 50\% of the peak flux density on either side of the profile. The 50\% flux density level is estimated using linear fits ($f=a+bv$) to each side of the profile between 15-85\% of the peak flux (minus the rms noise) on either side of the profile The width is corrected for instrumental broadening and cosmological stretch.\newline
    
    \item[$\boldsymbol{\Delta s}$:] Correction for linewidth overestimation due to instrumental broadening following \citet{Springob05}, defined as
    \begin{equation}
        \Delta s = \frac{2\Delta v \lambda}{1+z}
    \end{equation}
    This factor is already subtracted from the linewidths tabulated here, but is provided here in case the user wants to remove it and apply alternative corrections to the GBT data. We do not provide this value for ALFALFA data since \citet{Haynes18} only provides linewidths that are already corrected. \newline
    \item[$\boldsymbol{V_{\rm HI}}$: ]Centroid of the {\hi} line detection, calculated by taking the midpoint of the velocities at the 50\% peak flux density level on either side of the {\hi} profile determined when calculating $W_{F50}$.\newline
    \item[$\boldsymbol{P_r}$:] The peak {\hi} flux at the high velocity peak (identical to $P_l$ for a single-peaked profile). \newline
    \item[$\boldsymbol{P_l}$:] The peak {\hi} flux in the low velocity peak (identical to $P_r$ for a single-peaked profile).\newline
    \item[$\boldsymbol{a_r}$:] $y$-intercept fit parameter for high velocity side of the {\hi} profile, used to calculate $W_{F50}$.\newline
    \item[$\boldsymbol{b_r}$:] slope fit parameter for high velocity side of the {\hi} profile, used to calculate $W_{F50}$.\newline
    \item[$\boldsymbol{a_l}$:] $y$-intercept fit parameter for low velocity side of the {\hi} profile, used to calculate $W_{F50}$.\newline
    \item[$\boldsymbol{b_l}$:] slope fit parameter for low velocity side of the {\hi} profile, used to calculate $W_{F50}$.\newline
    \item[$\boldsymbol{\sigma_{V_{\hi}}}$] Error on $v_{\hi}$, defined as
    \begin{equation}
        \sigma_{V_{HI}} = \frac{1}{2}\sqrt{\left(\frac{rms}{b_l}\right)^2 + \left(\frac{rms}{b_r}\right)^2}
    \end{equation}
    \newline
    \item[\bf confusion flag:] Flag to indicate possible confusion of the {\hi} profile (0$=$not confused, 1$=$possibly confused).  \newline
    \item[$\boldsymbol{P_{R>0.2}}$:] Probability that more than 20\% of the total measured flux comes from galaxies other than the primary. See Section~\ref{sec:confusion} for details.  \newline
    \item[\bf OFF detection flag:] Flag to indicate the presence of an {\hi} source in the OFF observation which has a similar velocity as the redshift of our target. The presence of {\hi} emission in the OFF beam may lead to an underestimation of the flux from our primary target.\newline
    \item[\bf Baseline structure flag]: Flag indicating the presence of strong baseline variations which limit our ability to measure {\hi} source properties. Such structure can hamper fitting a smooth baseline to the data, and may have structure on scales similar to typical galaxy linewidths, making differentiating baseline structure from true signal challenging.
    
\end{enumerate}
Any missing measurements (e.g., linewidths for HI non-detections) are represented by \texttt{-999} in the catalog. All linewidths include a redshift stretching correction factor of $(1+z)^{-1}$, as well as an  instrumental broadening correction following \citet{Springob05}. We have not applied corrections for turbulent motion or galaxy inclination due to their potentially large uncertainties, but for convenience we have included inclination estimates in our catalog. Multiple linewidth estimates are provided to enable comparison with other studies, although we consider $W_{F50}$ the most reliable because it is the least sensitive to low $S/N$ data at the edges of {\hi} profiles \citep{Giovanelli97,Springob05}.  However, all linewidths can become unreliable at $S/N\lesssim5$ due to a combination of both the random noise coupled with the difficulty identifying peaks in {\hi} profiles \citep{Stark13}.  

\begin{figure}
    \centering
    \includegraphics[width=\columnwidth]{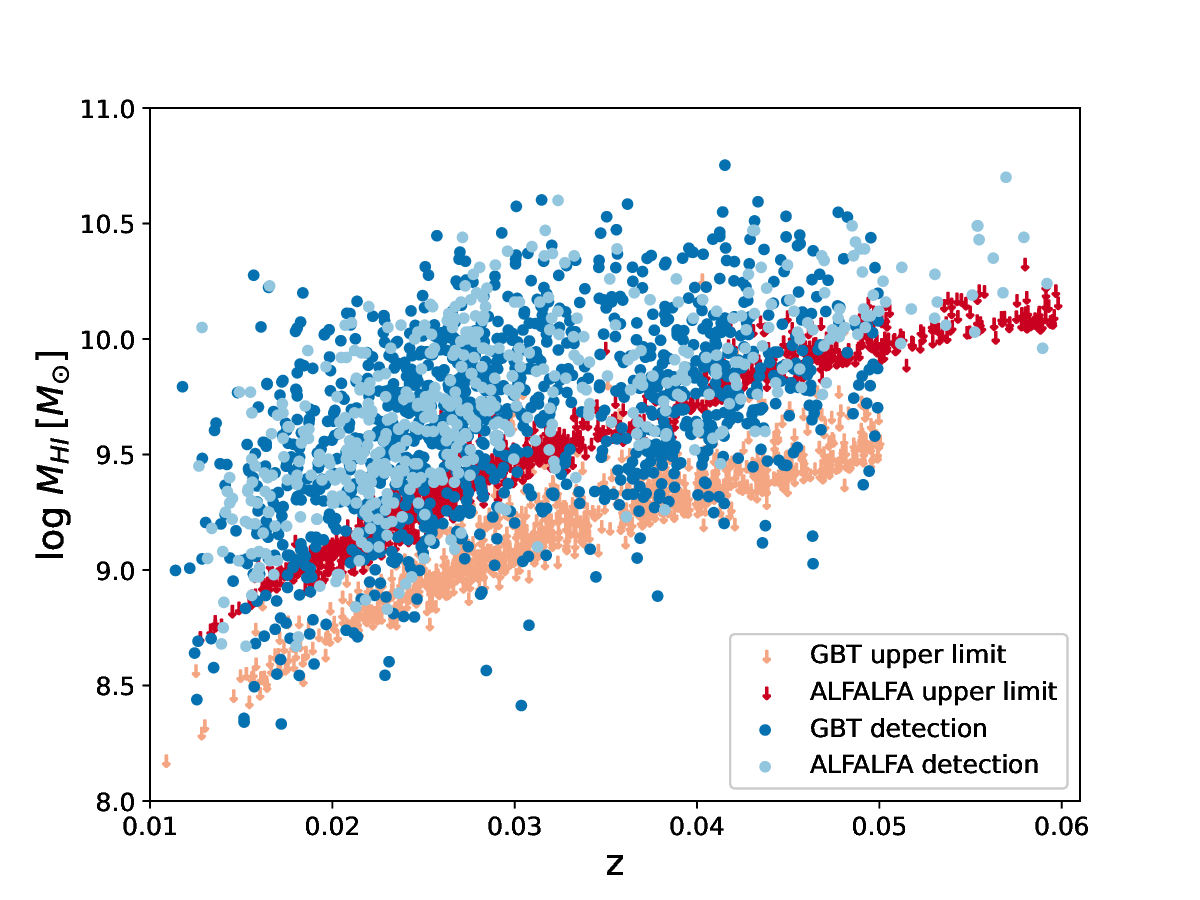}
    \caption{Derived {\hi} mass versus redshift for the HI-MaNGA DR2 catalog. Upper limits from ALFALFA are 4.5$\sigma$ while GBT upper limits are 3$\sigma$. The more conservative upper limits for ALFALFA data are due to the imposed minimum $S/N$ in the a.100 catalog of detections.}
    \label{fig:mhi_vs_z}
\end{figure}
 
\begin{figure*}
    \centering
    \includegraphics[width=2\columnwidth]{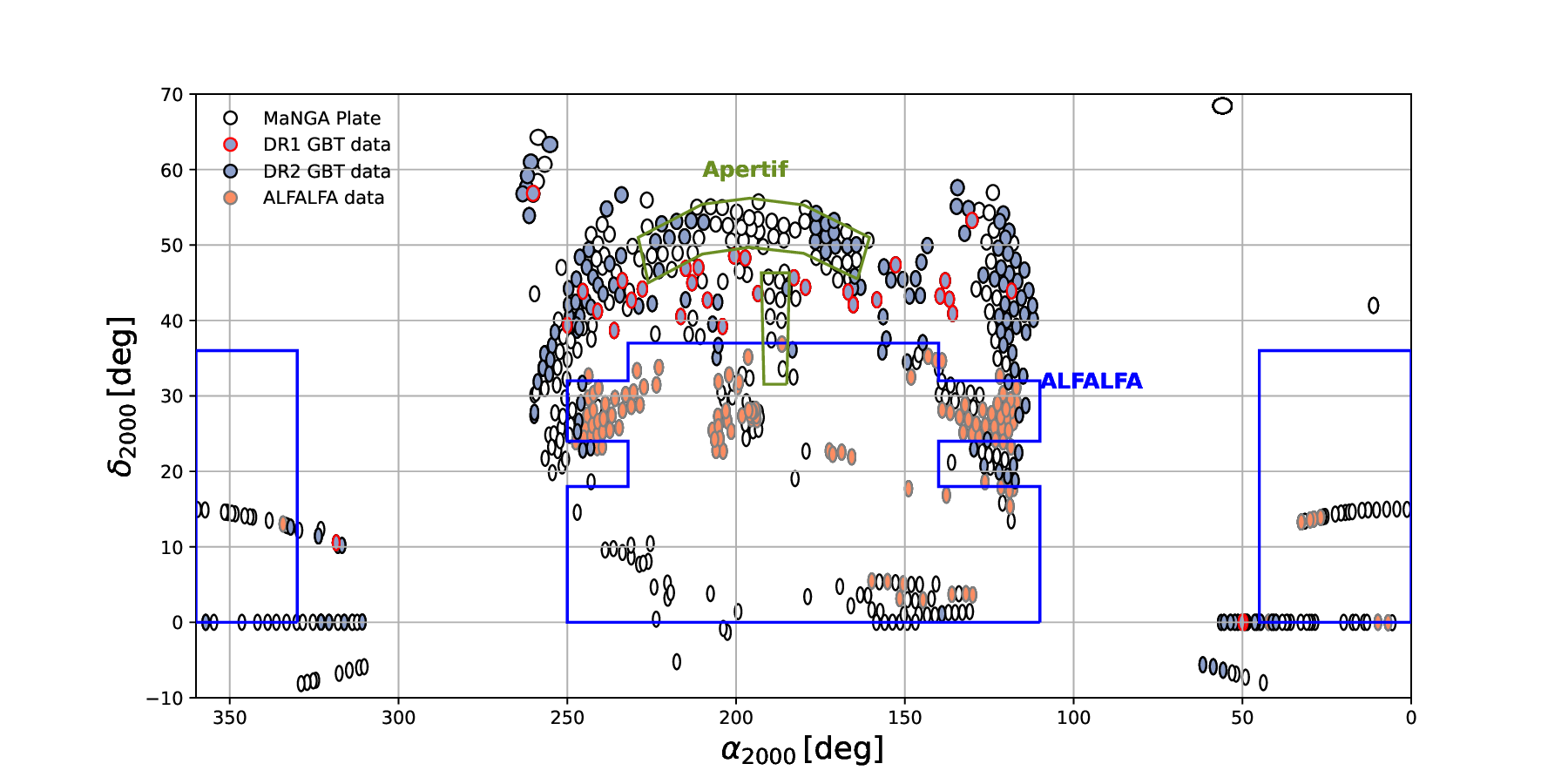}
    \caption{Distribution of plates for the complete MaNGA survey and those with follow-up {\hi} observations released in DR2. The footprints of ALFALFA and the expected Apertif Medium Deep surveys are overlaid. Plates in the ALFALFA footprint not currently highlighted as having ALFALFA data will have their ALFALFA crossmatch released in the future.}
    \label{fig:sky_cov}
\end{figure*}

Figure~\ref{fig:mhi_vs_z} shows the derived {\hi} mass as a function of redshift for our catalog, including both new GBT data and the ALFALFA crossmatch. Fig~\ref{fig:sky_cov} shows the sky coverage of HI-MaNGA observations.

We also make available \texttt{fits} and \texttt{csv} files containing the spectrum of each galaxy. These files contain the following columns:
\begin{enumerate}
    \item [$\boldsymbol{v_{HI}}$:] Barycentric recession velocity for each channel (optical convention). \newline
    \item[$\boldsymbol{F_{HI}}$:] Channel flux density after baseline subtraction.\newline
    \item[$\boldsymbol{B_{HI}}$:] Channel flux density prior to baseline subtraction; provided so a user can subtract their own baseline if desired.
\end{enumerate}
 
 \subsection{Analysis Sample}
 \label{sec:analysis_sample}

Our parent sample is from the internal MaNGA Product Launch 8 (MPL-8), which is composed of 6583 unique galaxies with a roughly uniform stellar mass distribution from $\log{M_*/M_{\odot}}=8.5-11.5$ (Chabrier Initial Mass Function, IMF; \citealt{Chabrier03}). When crossmatched with the HI-MaNGA catalog, we are left with 3669 galaxies with {\hi} data.  The MaNGA survey is designed such that more massive galaxies are chosen to be at larger distances, ensuring they can have spectroscopic coverage out to the same relative radius as the rest of the sample. This distance-mass dependence, combined with the fact that HI-MaNGA only observes out to $z\sim0.05$, means the data set with {\hi} is somewhat biased against the most massive galaxies ($\log{M_*/M_{\odot}}>11$; see Fig.~1 of \citealt{Masters19}).

For galaxies which have data from both GBT and ALFALFA, we use the following criteria to determine which observation to use. If both observations are detections, and neither are confused, we use the higher $S/N$ observation. If one of the observations is confused, we use the non-confused observation.  A case where both observations are confused is irrelevant because these are not included in any analysis.  If both observations are nondetections, we use the observation with the lowest rms noise level. If one observation is a detection and the other a non-detection, we use the detection as long as it is not flagged as confused. Otherwise, the upper limit is used. Any galaxies with observational artifacts which may impact the \ion{H}{i} mass measurement, such as strong baseline oscillations or ``negative" detections caused by a galaxy in the OFF beam at similar redshift as the primary target, are rejected from our analysis.

We limit our analysis sample to $z<0.05$, removing any ALFALFA galaxies with slightly higher redshift. We use the Primary+ MaNGA subsample for our analysis, including the Primary sample with a flat stellar mass selection, as well as the ``color-enhanced" sample designed to more evenly incorporate less-populated regions in color-stellar mass parameter space \citep{Wake17}. The IFU bundles measure spectra out to $\sim1.5R_e$ in this subsample, where $R_e$ is the $r$-band elliptical Petrosian half-light radius.  We do not include the Secondary sample, with flat stellar mass selection and IFU coverage out to $\sim2.5R_e$, as it lies at larger distance and has a substantially higher {\hi} non-detection rate.

Our analysis focuses only on star forming galaxies, specifically any galaxy which falls in the star-forming region of both the [\ion{O}{iii}]/H$\beta$ vs. [\ion{N}{ii}]/H$\alpha$ and [\ion{O}{iii}]/H$\beta$ vs. [\ion{S}{ii}]/H$\alpha$ Baldwin, Phillips \& Telervich \citep[BPT][]{Baldwin81} diagrams of \citet{Kewley06b}, where the line ratios are integrated out to $R_e$. We further require galaxies to have integrated H$\alpha$ equivalent widths, EW(H$\alpha$), measured over the same area, of $>6${\AA}, placing them on the so-called ``star forming sequence". There are an additional 20 galaxies which fall in the star-forming region on the BPT diagrams but have $3<{\rm EW}(H\alpha)<6$, making them ``green valley" galaxies possibly transitioning between the star forming and passive populations \citep{Sanchez14,Cano-Diaz19}. We do not include this subsample in any statistical analysis although these data are shown in figures throughout this paper to illustrate how relationships between {\hi} content and optical ISM diagnostics may evolve as galaxies migrate off the star-forming sequence. To avoid active galactic nuclei (AGN), including those embedded in star-forming disks, we also require that the line ratios measured within the central 2.5{\arcsec} fall within the star-formation region of the BPT diagrams. Future work will explore {\gs} in AGN and passive galaxies. 

\begin{figure}
     \centering
     \includegraphics[width=\columnwidth]{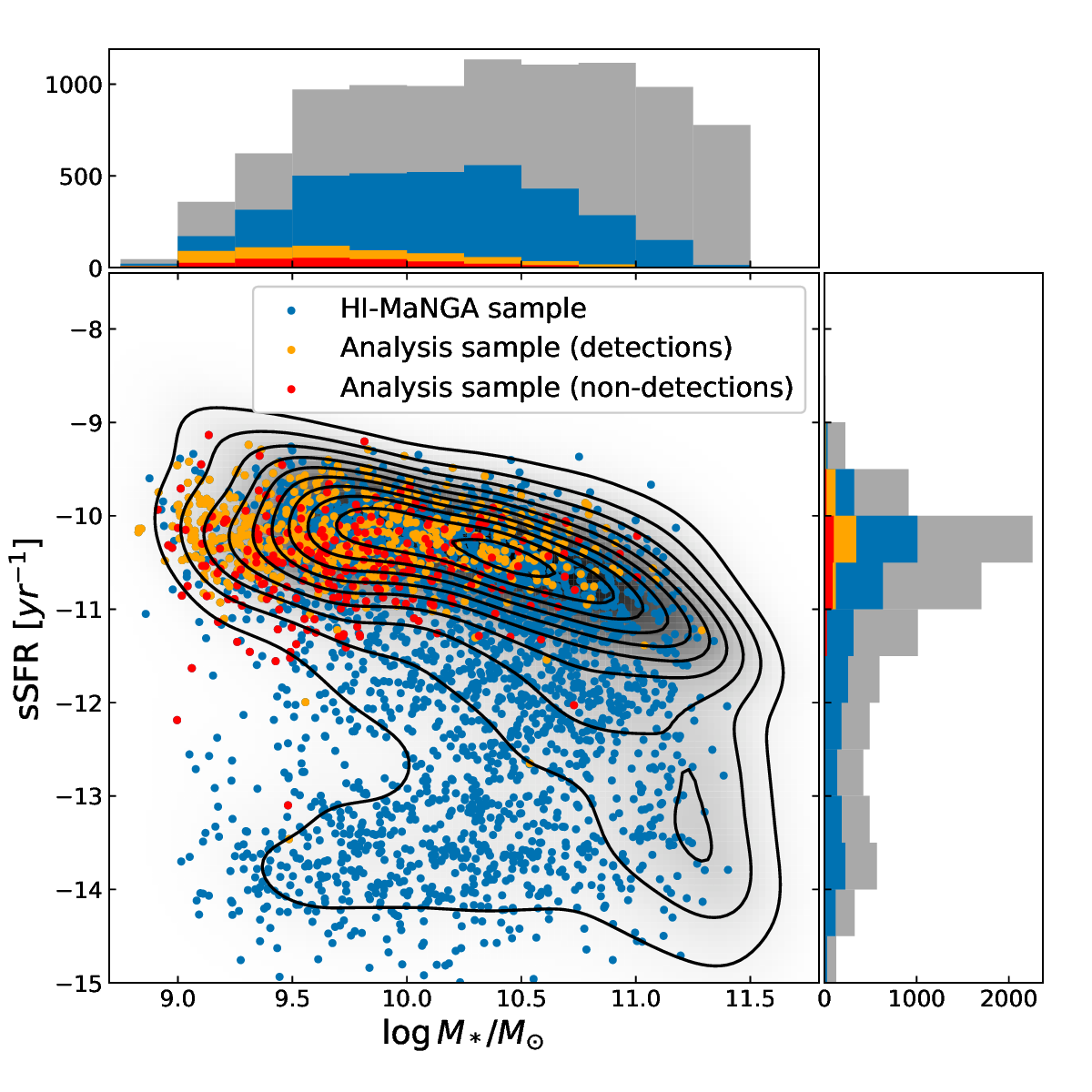}
     \caption{The HI-MaNGA sample and the analysis sample for this paper in sSFR vs. $\log{M_*}$ parameter space. Black contours show the distribution for the parent MaNGA sample. SFRs taken from the \texttt{Pipe3d} analysis of MaNGA data cubes (see Section~\ref{sec:manga_properties}).}
     \label{fig:sample}
 \end{figure}

The following is a summary of each sample cut and the total number of objects removed from the starting sample of 3669 unique galaxies by each cut: $p_{R>0.1} < 0.1$ (494), No OFF detection flag or baseline structure flag (83), $z<0.05$ (199), MaNGA Primary+ sample only (1123), line ratios within $R_e$ consistent with star formation using BPT diagrams (1784), EW(H$\alpha$)$>6$ (1591), line ratios measured with central 2.5$\arcsec$ inconsistent with AGN using BPT diagrams (928). After these various selections, we are are left with a sample of 837 galaxies, including 607 detections (586 of which have $S/N > 5$) and 230 non-detections. These numbers do not include the 20 green valley galaxies described above. Figure~\ref{fig:sample} illustrates the parent MaNGA sample, the full HI-MaNGA catalog sample, and the sample used for our analysis in sSFR vs. $M_*$ space.
 
\subsection{MaNGA Data Products and ISM Diagnostics}
\label{sec:manga_properties}

Details of MaNGA instrumentation, observing strategy, data reduction pipeline (DRP), data analysis pipeline (DAP), and data products can be found in \citet{Drory15}, \citet{Law15},\citet{Law16}, \citet{Yan16b}, \citet{Westfall19}, \citet{Aguado19}, and \citet{Belfiore19}. Although our parent sample is taken from MPL-8, we use more up-to-date data products from MPL-9.  All measurements are derived from maps of each spectral quantity using the \texttt{SPX} binning scheme, where each $0.5\arcsec\times0.5\arcsec$ spaxel is analyzed independently.  Below we describe each of the parameters measured and their use in characterizing the ISM. Unless otherwise noted, spectral properties are all measured within $R_e$ in order to provide consistent measurements across our full sample. In cases where we measure line ratios, we measure the ratio spaxel-by-spaxel and then take the median within our region of interest. When measuring line ratios, we require all relevant lines to have $S/N>3$. Throughout our analysis, we use the stellar masses from the NSA catalog which are estimated using \texttt{kcorrect} with Galaxy Evolution Explorer ($GALEX$) and SDSS photometry \citep{Blanton07b}. 

In part of our analysis, we use parameters measured from an analysis of MaNGA data using \texttt{Pipe3d}. We refer the reader to \citealt{Sanchez16a} and \citealt{Sanchez16b} for details. Specifically, we make use of two estimates of star formation rate, one measured from extinction-corrected H$\alpha$ flux and the other from stellar population synthesis modeling. A constant factor of $-0.24$ dex is subtracted from all \texttt{Pipe3d} SFRs to convert from a Salpeter IMF \citep{Salpeter55} to a Chabrier IMF used by the MaNGA DAP.

\subsubsection{Equivalent Width, EW}
Equivalent width measures emission line flux normalized by the strength of stellar continuum. When examining how {\gs} scales with the strength of individual emission lines, it is appropriate to normalize emission line fluxes by the stellar continuum level in a way that is analogous to how {\gs} normalizes the \ion{H}{i} mass by the stellar mass. 

The DAP measures equivalent widths for each bin, which in the case of the \texttt{SPX} binning scheme, is each 0.5\arcsec$\times$0.5\arcsec spaxel. The continuum flux level used in the calculation for each emission line and spaxel is recorded.  Therefore, the integrated equivalent width is calculated simply by dividing the summed emission line fluxes by the summed continuum flux densities within $R_e$. Our analysis considers all measured lines from 3727{\AA} ([\ion{O}{ii}]) to 9548{\AA} (P$\epsilon$).  

\subsubsection{Gas-Phase Metallicity, 12 + log O/H}
\label{sec:metallicity}
The metal abundance (metallicity) has a profound impact on the ISM, dramatically increasing the cooling rate and increasing the ability of clouds to self-shield from ionizing radiation. There are many calibrations between gas-phase metallicity and various strong emission line ratios.  Unfortunately, there are also significant disagreements in the absolute metallicities provided by different calibrations, although {\it relative} metallicity variations tend to be more consistent \citep{Kewley08}.  For this study, we primarily care about relative trends, and we use the N2O2 calibrator from \citet{Kewley02} which predicts the metallicity (12 + log O/H) as a function of the [\ion{N}{ii}]6585\AA/[\ion{O}{ii}]3727{\AA} ratio.  The strong advantage of this calibration is its insensitivity to other local ISM properties, notably the ionization parameter, $q$ (see Section~\ref{sec:ionization_parameter}), and contamination by emission from diffuse ionized gas, or DIG \citep{Zhang17}.  As will be discussed in Section~\ref{sec:role_of_dig}, variations in the scaling of {\gs} against EW may be associated with the fraction of emission from DIG, necessitating the need for a metallicity calibration that is insensitive to exactly where the line emission is originating.  A disadvantage of the N2O2 method is that it can only be used for metallicities above $\sim0.5Z/Z_{\odot}$.  However, we expect all metallicities in our sample to fall above this threshold.

Since [\ion{O}{ii}] 3727{\AA} and [\ion{N}{ii}] 6585{\AA} lie at significantly different wavelengths, we apply internal extinction corrections before measuring their ratio. We use the Balmer decrement to estimate $A_v$ assuming an intrinsic H$\alpha$/H$\beta$ ratio of 2.86 \citep{Osterbrock06} and the extinction curve of \citet{Odonnell94}. We require H$\alpha$ and H$\beta$ to be detected to $S/N>3$.

\subsubsection{Ionization Parameter, $q$}
\label{sec:ionization_parameter}
The ionization parameter is the maximum velocity of an ionization front driven by a local radiation field, typically defined as,
\begin{equation}
q=S_{H^0}/n_H
\end{equation}
where $S_{H^0}$ is the flux of ionizing photons per unit area and $n$ is the hydrogen number density, including all ionized and neutral hydrogen \citep{Kewley02}.  This parameter serves as a useful indicator of the ionization state of the ISM. We estimate $q$ as a function of [\ion{O}{iii}] 5008\AA/[\ion{O}{ii}] 3727,3729{\AA} and $12+\log{\rm O/H}$ using the calibration of \citet{Kewley02}. We apply extinction corrections to the [\ion{O}{iii}] and [\ion{O}{ii}] lines in the same manner as described in Section~\ref{sec:metallicity}.

\subsubsection{Electron density, $n_e$}
Electron density affects the collision rates of particles in the ISM, which can in turn affect the strengths of the forbidden lines used in our analysis.  The ratio [\ion{S}{ii}]6718{\AA}/[\ion{S}{ii}]6732{\AA} provides an estimate of electron density due to the fact that these lines are emitted by two different energy levels with very similar excitation energy, making their relative strengths depend primarily on the collision strength.  The ratio is sensitive to variations in $n_e$ between ${\sim}100$ and ${\sim}10^4$ cm$^{-3}$ corresponding to line ratios between ${\sim}1.4$ and  ${\sim}0.3$ \citep{Osterbrock06}.  

\subsubsection{[N\,{\sevensize II}]/H$\alpha$, [S\,{\sevensize II}]/H$\alpha$, [O\,{\sevensize I}]/H$\alpha$}
The [\ion{N}{ii}]6586\AA, [\ion{S}{ii}]6718,6732\AA, and [\ion{O}{i}]6302{\AA} lines are low ionization lines which are emitted in partially ionized regions of the ISM.  Around star-forming regions, these lines are emitted from the transition zone between the ionized and neutral gas at the edges of star-forming clouds. The strengths of these lines relative to $H\alpha$ are thought to be significantly enhanced in the presence of a harder ionizing field (relative to HII regions) and shocks, and they often coincide with regions ionized by AGN or massive evolved stars \citep{Veilleux87,Stasinska08,CidFernandes11,Belfiore16}.  These line ratios are also strengthened in DIG by factors of several relative to HII regions \citep{Reynolds85a, Reynolds85b, Reynolds98, Haffner99, Hoopes03, Madsen06,Voges06,Oey07,Zhang17}.   

\subsubsection{H$\alpha$ surface brightness, $\mu_{H{\alpha}}$}
While H$\alpha$ emission from HII regions is a known tracer of star formation \citep{Kennicutt12}, when measured over large scales it can be contaminated by emission from DIG residing beyond HII regions \citep{Oey07}.  The effective MaNGA point spread function (PSF; $\sim$2.5\arcsec) corresponds to spatial scales of $\sim$1 kpc, significantly larger than typical HII regions, so we cannot determine the fraction of H$\alpha$ emission arising from HII regions versus DIG in our data.  However,  H$\alpha$ surface brightness, $\mu_{\rm H\alpha}$ can be used as a an indicator of the contribution of DIG to the overall H$\alpha$ emission, as supported by observations which show increasing low-ionization line strength with decreasing $\mu_{\rm H\alpha}$  \citep{Oey07,Zhang17}.  We measure the median value of $\mu_{H\alpha}$ within $R_e$ to characterize the average impact of DIG in a relative sense throughout our sample. It is not corrected for internal reddening.

\subsection{Treatment of upper limits}
\label{sec:upper_limits}
Even the deepest \ion{H}{i} surveys do not detect all optically identified targets, but the \ion{H}{i} mass upper limits derived from non-detections still hold valuable information.  Ignoring non-detections when examining relationships between \ion{H}{i} mass and other galaxy properties can be misleading, causing artificially strong or weak correlations, depending on the exact sampling function \citep[e.g.][]{Calette18}.

HI-MaNGA detects approximately 55\% of all observed targets and 70\% in our analysis sample. In the analyses presented in this paper, we employ two distinct methods to study population trends in the presence of non-detections: (1) survival analysis with the generalized Kendall's $\tau$ and the Akritas-Theil-Sen Estimator to conduct correlation tests and perform linear fits in the presence of non-detections,  and (2) the Photometric Gas Fractions technique to replace upper limits with {\hi} mass estimates based on other galaxy properties.

\subsubsection{Survival analysis with the generalized Kendall's $\tau$ and Akritas-Theil-Sen Estimator}
\label{sec:ats}
 Making inferences from data in the presence of non-detections (or ``left censored" data) has been a pursuit of the specific branch of statistics known as ``survival analysis", and there exist a number of studies where these methods have been applied to astronomical data \citep{Feigelson85, Schmitt85, Isobe86, Akritas95, Akritas96, Feigelson12, Calette18, Yesuf19,Rodriguez-Puebla20}. Our goal is to test the strengths and statistical significances of correlations between {\gs} and other parameters, while also determining the best linear relations, all with a significant fraction of {\hi} non-detections. To achieve this goal, we follow the methodology outlined by \citet{Akritas95}. First, to test correlation strengths and significances, we employ a generalized version of Kendall's $\tau$ correlation test which incorporates upper limits.  Building directly on the generalized Kendall's $\tau$ is the Theil-Sen estimator to determine a best-fit line (as \citealt{Akritas95} developed the generalized Theil-Sen estimator used here, we refer to the method as the Akritas-Theil-Sen estimator, or ATS).  This method determines the slope, $a$, between the response variable, $y$, and the covariate, $x$, by finding the value of $a$ where $\tau$ calculated between the residuals, $y-ax$, and $x$ is approximately zero.  The $y$-intercept is then found by estimating the median of $y-ax$. To find the median when the residuals contain upper limits, we use the Kaplan-Meier estimator \citep{Kaplan58} to estimate their survival function, $S(x)$.  The survival function provides the likelihood that a distribution has a value above $x$, and relates to the cumulative distribution function, $F(x)$, by $S(x) = 1-F(x)$.  The intercept is thus where the Kaplan-Meier estimator equals 0.5.  The Kaplan-Meier estimator of the fit residuals also allows us to estimate the 1-$\sigma$ scatter around the mean relation, which is done by finding where it equals 0.16 and 0.84
 
In practice we use the \texttt{R} packages \texttt{cenken} and \texttt{survfit} to conduct these analysis. To quantify uncertainties on the estimated slopes and intercepts, we use bootstrapping following the recommendation of \citet{Wilcox10}. We resample $N$ pairs of data points from the $N$ original pairs (preserving whether each pair was an {\hi} detection or upper limit) and re-estimate the fit parameters.  This process is repeated $600$ times and the middle 68\% of the returned distributions are used to define the confidence intervals on our fit parameters. 

The ATS estimator as applied here assumes all relationships are linear. This assumption generally appears to be valid, although there may be some exceptions, notably analyses involving ionization parameter, $q$. Kendall's $\tau$ assumes a monotonic relationship between variables, not necessarily a linear one, so should be a generally reliable indicator of correlation strength. 

To test that the ATS estimator gives reliable and consistent linear fits in the presence of upper limits, we have conducted a series of fits between {\gs} and $g-r$ color for a mock data set where we increase the fraction of censored data in a manner consistent with observing \ion{H}{i} in MaNGA galaxies to progressively shallower depth.  We find the linear fit parameters returned by the ATS estimator are very robust to variations in the fraction of censored data and consistently agree well with the ``true" linear fit to the uncensored data, whether it is determined using the Theil-Sen estimator or a more traditional least-squares approach, as long as the censoring fraction does not exceed $\sim50\%$. Once censored data dominate the sample, the ATS estimator results fits that are biased, although not as biased as a na\"ive fit to only detections. This test is described further in Appendix~\ref{sec:app_ast}.

We have also examined whether localized regions of parameter space that are dominated by non-detections can bias the resulting ATS fit parameters. To assess this possibility, we test whether our results in Section~\ref{sec:results} are sensitive to whether the ATS estimator is applied to data with and without regions of parameter space dominated by upper limits. We find no significant change to our results, but out of an abundance of caution we still limit our fits to regions dominated by {\hi} detections. Further details are provided in Section~\ref{sec:results}. 

The tests described here are by no means an exhaustive analysis of the limitations of the ATS estimator, but do assess its applicability to our own data set, and we find that it should perform reliably for our purposes. The ATS estimator is the primary tool used to conduct statistical analysis in the presence of {\hi} non-detections throughout this work.

\subsubsection{Photometric Gas Fractions}
\label{sec:photogs}
 An alternative approach to handing {\hi} upper limits is to replace them with estimates of {\hi} content using other galaxy properties. In Section~\ref{sec:photogs_compare} we examine how our results may change using this approach as opposed to using the ATS estimator. We follow a method very similar to that described in \citet{Eckert15} to assign gas masses to our non-detections.  Using data from the REsolved Spectroscopy Of a Local VolumE (RESOLVE) survey\footnote{\url{https://resolve.astro.unc.edu}}, \citet{Eckert15} fit a model to the 2D probability distribution of galaxies in the {\gs} vs. modified color plane ($P(M_{\ion{H}{i}}/M_*|\,{\rm color})$, where modified color refers to a linear combination of color with another galaxy parameter. The model is composed of two major components: {\hi} detections which follow a linear relationship with modified color with some scatter, and non-detections which begin to appear at larger modified colors and cluster around \gs$=0.05$. From $P(M_{\ion{H}{i}}/M_*|\,{\rm color})$, upper limits can be assigned new estimates of {\gs} as long as their modified color is measured. However, MaNGA photometry comes from the NSA, and there are likely systematic differences between NSA and RESOLVE photometry, so we do not use the same modified color and fitted model parameters from \citet{Eckert15} to predict {\gs} for non-detections. Instead, we crossmatch the RESOLVE survey with the NSA, then rerun the analysis of \citep{Eckert15} to determine $P(M_{\ion{H}{i}}/M_*|\,{\rm color})$. The full details of our analysis are given in Appendix~\ref{app:photogs}.  

\section{Results}
\label{sec:results}
We now examine how {\gs} is correlated with different parameters derived from MaNGA optical spectroscopy.  In Section~\ref{sec:gs_ew} we characterize scaling relationships between {\gs} and emission line equivalent widths, followed by testing correlations between {\gs} and other ISM diagnostics in Section~\ref{sec:gs_ism_properties_corr}.  In Section~\ref{sec:resid}, we examine whether the scatter in the {\gs}-EW relations can be attributed to variations in their typical ISM properties.

\subsection{$\text{M}_\text{H{\sc i}}/\text{M}_*$ vs. EW relations}
\label{sec:gs_ew}

\begin{figure*}
    \centering
    \includegraphics[width=\columnwidth]{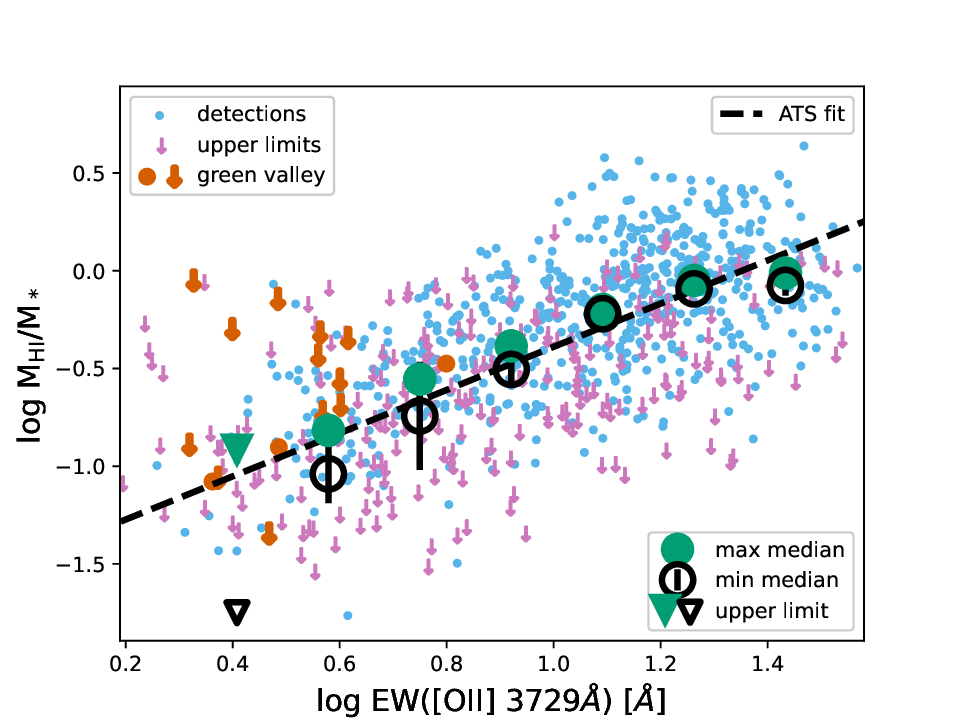}
    \includegraphics[width=\columnwidth]{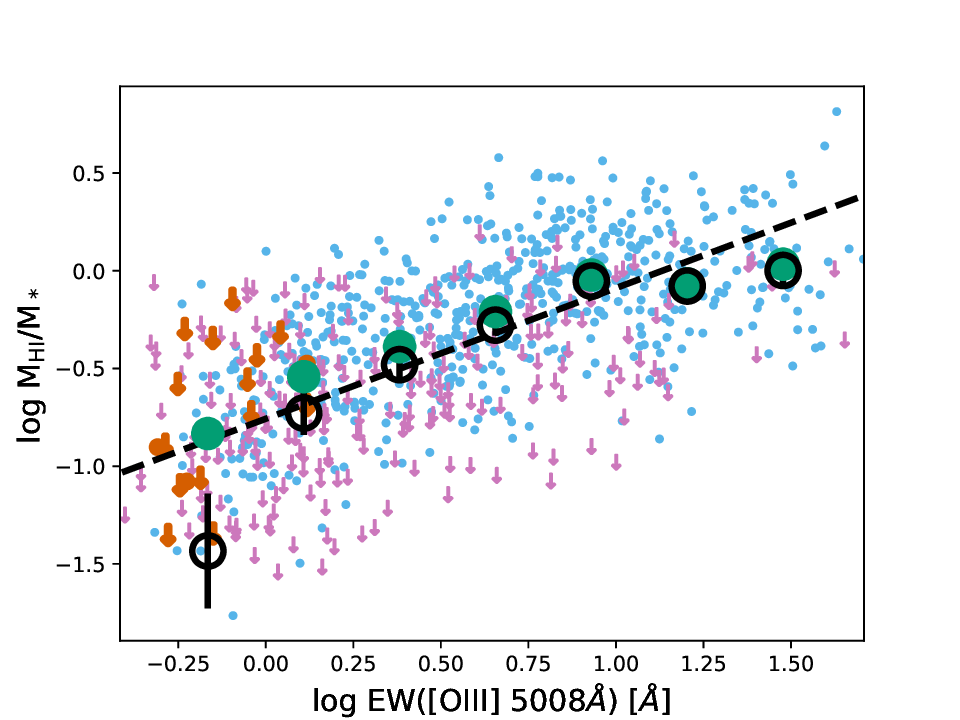}
    \includegraphics[width=\columnwidth]{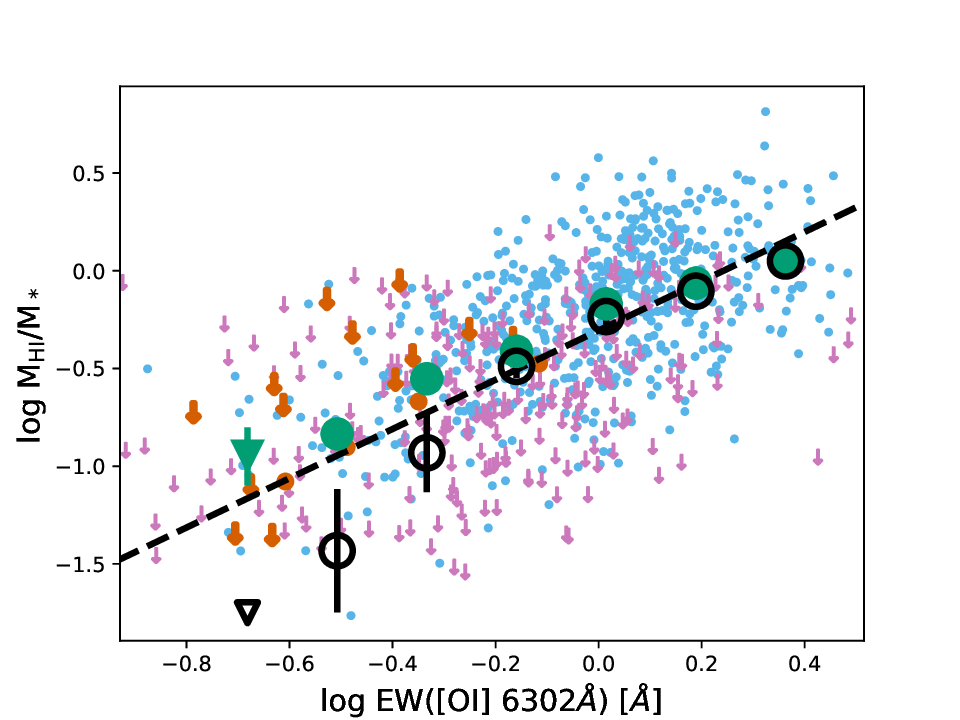}
    \includegraphics[width=\columnwidth]{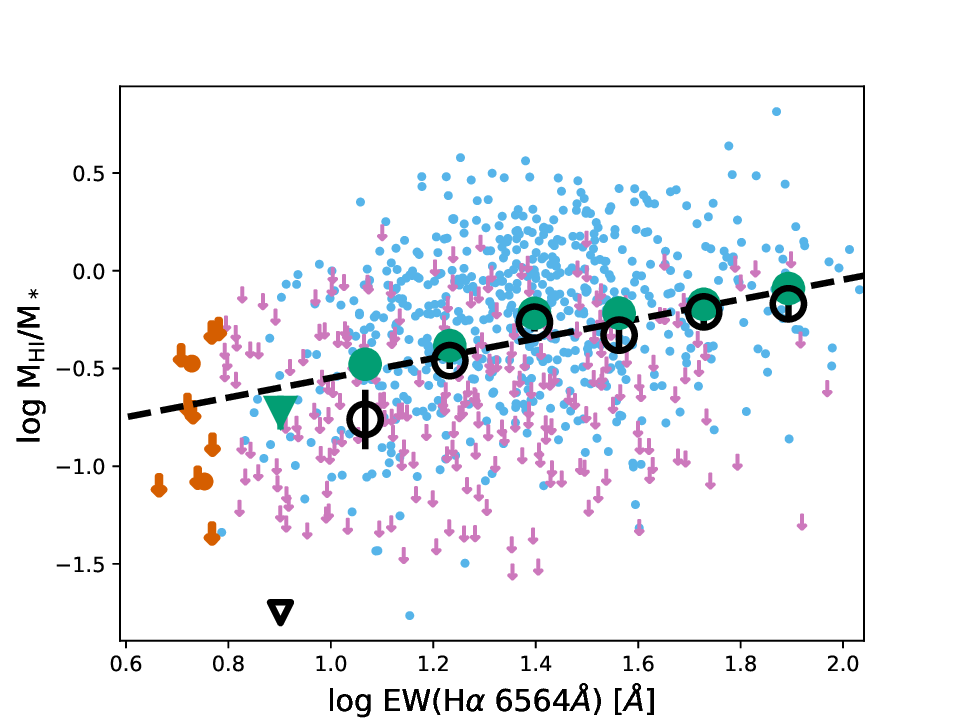}
    \caption{$EW$ vs. {\gs} for a subset of optical emission lines analyzed in this work.  The three oxygen lines ([\ion{O}{i}],[\ion{O}{ii}],[\ion{O}{iii}]) show the strongest correlations among all emission lines.  $H\alpha$ (bottom-right) is significantly weaker. Black lines show the linear fit derived from the Akritas-Theil-Sen estimator which accounts for upper limits (see Section~\ref{sec:ats}). Large green filled and black unfilled points represent binned medians under two different assumptions about the upper limits: either the true {\gs} is equal to the measured upper limit, or it is zero. Inverted triangles represent bins which have at least 50\% non-detections, such that their minimum binned median is zero. Thick orange points represent galaxies in the green valley.}
    \label{fig:ew_gs}
\end{figure*}

We begin by simply testing the correlation strengths between all measured optical emission lines and {\gs}. Instead of using emission line luminosities, we use emission line equivalent widths (EWs), which normalize the emission line strength by the stellar continuum strength in a manner analogous to the gas-to-stellar mass ratio. Table~\ref{tab:scaling_tau} summarizes the correlations between each emission line EW and {\gs}. The table includes the generalized Kendall's $\tau$ correlation coefficient and the corresponding p-value, the linear fit coefficients and their uncertainties determined using the ATS estimator, and the scatter above and below the best-fit line ($\sigma_h$, $\sigma_l$).  

\begin{table*}
    \caption{{\gs} correlation statistics}
    \centering
    \begin{threeparttable}
    \begin{tabular}{lllllll}
    \hline
    covariate & $\tau$ & p-value$^*$ & slope & intercept & $\sigma_h^{\dagger}$ & $\sigma_l^{\dagger}$\\
    \hline
     \multicolumn{6}{c}{Equivalent Widths} \\
     \hline
     $[\ion{O}{ii}]$ 3727\AA &   0.37 &   $<$\num{2.22E-16} &   1.08$\pm$  0.05 &  -1.33$\pm$  0.05 &   0.35 &   0.50 \\
$[\ion{O}{ii}]$ 3729\AA &   0.38 &   $<$\num{2.22E-16} &   1.11$\pm$  0.05 &  -1.49$\pm$  0.06 &   0.34 &   0.48 \\
H12 3751\AA &   0.21 &   $<$\num{2.22E-16} &   0.83$\pm$  0.08 &  -0.11$\pm$  0.03 &   0.40 &   0.67 \\
HII 3771\AA &   0.22 &   $<$\num{2.22E-16} &   0.79$\pm$  0.07 &  -0.04$\pm$  0.03 &   0.39 &   0.68 \\
H$\theta$ 3798\AA &   0.20 &   $<$\num{2.22E-16} &   0.89$\pm$  0.10 &  -0.09$\pm$  0.03 &   0.41 &   0.67 \\
H$\eta$ 3836\AA &   0.25 &   $<$\num{2.22E-16} &   1.04$\pm$  0.09 &  -0.13$\pm$  0.03 &   0.42 &   0.59 \\
$[\ion{Ne}{iii}]$ 3869\AA &   0.35 &   $<$\num{2.22E-16} &   1.10$\pm$  0.06 &  -0.19$\pm$  0.02 &   0.35 &   0.52 \\
$[\ion{He}{i}]$ 3889\AA &   0.22 &   $<$\num{2.22E-16} &   0.95$\pm$  0.10 &  -0.19$\pm$  0.02 &   0.42 &   0.60 \\
H$\zeta$ 3890\AA &   0.26 &   $<$\num{2.22E-16} &   1.17$\pm$  0.10 &  -0.02$\pm$  0.03 &   0.41 &   0.57 \\
$[\ion{Ne}{iii}]$ 3968\AA &   0.36 &   $<$\num{2.22E-16} &   1.09$\pm$  0.06 &   0.40$\pm$  0.05 &   0.34 &   0.51 \\
H$\epsilon$ 3971\AA &   0.23 &   $<$\num{2.22E-16} &   1.25$\pm$  0.12 &  -0.38$\pm$  0.02 &   0.42 &   0.57 \\
H$\delta$ 4102\AA &   0.21 &   $<$\num{2.22E-16} &   0.96$\pm$  0.10 &  -0.48$\pm$  0.03 &   0.41 &   0.56 \\
H$\gamma$ 4341\AA &   0.20 &   $<$\num{2.22E-16} &   0.87$\pm$  0.10 &  -0.73$\pm$  0.05 &   0.41 &   0.61 \\
$[\ion{He}{ii}]$ 4687\AA &   0.13 & \num{1.14E-08}  &   0.55$\pm$  0.09 &   0.07$\pm$  0.08 &   0.43 &   0.66 \\
H$\beta$ 4862\AA &   0.18 & \num{8.06E-14}  &   0.72$\pm$  0.09 &  -0.89$\pm$  0.08 &   0.42 &   0.59 \\
$[\ion{O}{iii}]$ 4960\AA &   0.40 &   $<$\num{2.22E-16} &   0.67$\pm$  0.03 &  -0.45$\pm$  0.02 &   0.34 &   0.49 \\
$[\ion{O}{iii}]$ 5008\AA &   0.40 &   $<$\num{2.22E-16} &   0.67$\pm$  0.03 &  -0.76$\pm$  0.03 &   0.34 &   0.48 \\
$[\ion{N}{i}]$ 5199\AA &   0.13 & \num{3.98E-08}  &   0.69$\pm$  0.12 &   0.08$\pm$  0.08 &   0.44 &   0.68 \\
$[\ion{N}{i}]$ 5201\AA &   0.02 & \num{3.65E-01}  &   0.13$\pm$  0.14 &  -0.30$\pm$  0.08 &   0.44 &   0.69 \\
$[\ion{He}{i}]$ 5877\AA &   0.25 &   $<$\num{2.22E-16} &   0.93$\pm$  0.08 &  -0.25$\pm$  0.03 &   0.37 &   0.57 \\
$[\ion{O}{i}]$ 6302\AA &   0.36 &   $<$\num{2.22E-16} &   1.26$\pm$  0.07 &  -0.31$\pm$  0.02 &   0.37 &   0.49 \\
$[\ion{O}{i}]$ 6365\AA &   0.36 &   $<$\num{2.22E-16} &   1.25$\pm$  0.06 &   0.31$\pm$  0.04 &   0.36 &   0.49 \\
$[\ion{N}{ii}]$ 6549\AA &  -0.14 & \num{3.23E-09}  &  -0.54$\pm$  0.09 &  -0.18$\pm$  0.04 &   0.38 &   0.76 \\
H$\alpha$ 6564\AA &   0.13 & \num{3.11E-08}  &   0.50$\pm$  0.09 &  -1.05$\pm$  0.13 &   0.43 &   0.63 \\
$[\ion{N}{ii}]$ 6585\AA &  -0.14 & \num{4.55E-09}  &  -0.54$\pm$  0.07 &   0.08$\pm$  0.06 &   0.38 &   0.76 \\
$[\ion{S}{ii}]$ 6718\AA &   0.22 &   $<$\num{2.22E-16} &   0.88$\pm$  0.07 &  -0.98$\pm$  0.06 &   0.39 &   0.59 \\
$[\ion{S}{ii}]$ 6732\AA &   0.23 &   $<$\num{2.22E-16} &   0.90$\pm$  0.08 &  -0.86$\pm$  0.05 &   0.39 &   0.60 \\
$[\ion{He}{i}]$ 7067\AA &   0.24 &   $<$\num{2.22E-16} &   0.74$\pm$  0.06 &   0.10$\pm$  0.04 &   0.41 &   0.58 \\
$[\ion{Ar}{iii}]$ 7137\AA &   0.33 &   $<$\num{2.22E-16} &   0.95$\pm$  0.05 &  -0.13$\pm$  0.02 &   0.38 &   0.51 \\
$[\ion{Ar}{iii}]$ 7753\AA &   0.30 &   $<$\num{2.22E-16} &   0.69$\pm$  0.04 &  -0.02$\pm$  0.02 &   0.38 &   0.55 \\
P$\eta$ 9017\AA &   0.08 & \num{1.18E-03}  &   0.31$\pm$  0.09 &  -0.26$\pm$  0.04 &   0.44 &   0.65 \\
$[\ion{S}{iii}]$ 9071\AA &   0.22 &   $<$\num{2.22E-16} &   0.60$\pm$  0.06 &  -0.52$\pm$  0.03 &   0.40 &   0.57 \\
P$\zeta$ 9231\AA &   0.24 &   $<$\num{2.22E-16} &   0.57$\pm$  0.05 &  -0.34$\pm$  0.02 &   0.41 &   0.62 \\
$[\ion{S}{iii}]$ 9231\AA &   0.24 &   $<$\num{2.22E-16} &   0.61$\pm$  0.06 &  -0.79$\pm$  0.05 &   0.40 &   0.56 \\
P$\epsilon$ 9548\AA &   0.13 & \num{5.87E-08}  &   0.36$\pm$  0.06 &  -0.34$\pm$  0.03 &   0.45 &   0.64 \\
\hline
\multicolumn{6}{c}{ISM Diagnostics}\\
\hline
$[\ion{N}{ii}]/H\alpha$ &  -0.45 &   $<$\num{2.22E-16} &  -1.89$\pm$  0.08 &  -1.48$\pm$  0.06 &   0.32 &   0.42 \\
$[\ion{S}{ii}]/H\alpha$ &   0.19 & \num{2.22E-16}  &   1.94$\pm$  0.22 &   0.43$\pm$  0.09 &   0.40 &   0.65 \\
$[\ion{O}{i}]/H\alpha$ &   0.32 &   $<$\num{2.22E-16} &   1.70$\pm$  0.12 &   2.02$\pm$  0.17 &   0.35 &   0.58 \\
$[\ion{O}{iii}]/H\beta$ &   0.42 &   $<$\num{2.22E-16} &   1.10$\pm$  0.05 &  -0.15$\pm$  0.02 &   0.32 &   0.51 \\
$[\ion{S}{ii}]6732$\AA/[\ion{S}{ii}]6718$\AA$ &   0.09 & \num{3.20E-04}  &   1.02$\pm$  0.26 &  -1.84$\pm$  0.37 &   0.46 &   0.70 \\
$q$ &  -0.04 & \num{1.19E-01}  &  -0.29$\pm$  0.18 &   1.83$\pm$  1.36 &   0.42 &   0.58 \\
$\mu_{H\alpha}$ &  -0.12 & \num{5.63E-07}  &  -0.22$\pm$  0.04 &   8.13$\pm$  1.75 &   0.43 &   0.70 \\
$12+\log{O/H}$ (N2O2) &  -0.46 &   $<$\num{2.22E-16} &  -2.66$\pm$  0.09 &  23.63$\pm$  0.78 &   0.31 &   0.39 \\   
    \end{tabular}
    \end{threeparttable}
    \begin{tablenotes}
    \item $^*$ \mbox{Values of $<2.22\times 10^{-16}$ represent the limits of numerical precision in the \texttt{R} code used for this statistical test.}
    \item $^\dagger$ These refer to the scatter above and below the fitted line.
    \end{tablenotes}
    \label{tab:scaling_tau}
\end{table*}

As discussed in Section~\ref{sec:ats}, we limit our statistical analysis to avoid regions of parameter space dominated by upper limits. These regions are identified by first binning along the $x$-axis variable using 7 bins centered on the median $x$ value and spanning the inner 98\% of all data. Any bin with a non-detection rate of more than 50\% is excluded from the ATS estimator. We also only include bins with at least 10 data points to avoid regions of parameter space which are poorly sampled. In the vast majority of cases, excluding these data have little impact on our results, but we nonetheless adopt this approach out of caution. This fitting approach is adopted throughout our paper.

All lines, with the exception of [\ion{N}{i}]5201{\AA} have statistically significant correlations with {\gs}, although not all correlations are particularly strong. In all cases, the scatter around the best-fit line is typically asymmetric, skewing towards larger values, by as much as a factor of $\sim$1.5, below the fitted lines. The scatter (here and in subsequent sections) also appears intrinsic rather than driven by observational uncertainty in M$_{\hi}$; we have rerun our analysis setting the minimum required $S/N$ at 3, 5, and 10, and the scatter decreases by $\sim5\%$ at most. Alternatively, some of the scatter may be driven by the significantly different spatial scales over which optical and 21cm line emission is measured in our sample. In Section~\ref{sec:radial_dep}, we explore this issue further.

Notably, {\gs} is very weakly correlated ($\tau=0.13$) with EW(H$\alpha$), despite the fact that both $EW(H\alpha)$ and {\gs} are known to relate to sSFR (this is further discussed in Section~\ref{sec:gs_halpha_why}). In contrast, the equivalent widths of the Oxygen forbidden lines ([\ion{O}{i}], [\ion{O}{ii}], and [\ion{O}{iii}]) all show the strongest correlation coefficients, with $\tau=0.36-0.4$ (when generically referring to all Oxygen line equivalent widths, we will hereafter use the shorthand EW(O)). The similar correlation strengths are observed despite these emission lines typically originating in different regions of the ISM with different conditions.  We explore the reason for the strong \gs-EW(O) relation in Section~\ref{sec:gs_oxygen_why}. Figure~\ref{fig:ew_gs} illustrates the observed scaling relations between {\gs} and equivalent widths of $H\alpha$, [\ion{O}{i}], [\ion{O}{ii}], and [\ion{O}{iii}], with binned medians and the best fit line using the ATS estimator overlaid.

{\gs} also shows a strong correlation ($\tau=0.35$) with [\ion{Ne}{iii}]3869/3968{\AA} equivalent width. The [\ion{Ne}{iii}] and [\ion{O}{iii}] lines are already known to strongly correlate due to their similar ionization structure and constant Ne/O abundances \citep{Perez-Montero07}. Thus, since EW([\ion{O}{iii}]  correlates well with {\gs}, EW([\ion{Ne}{iii}]) is almost guaranteed to as well.

As discussed in Section~\ref{sec:analysis_sample}, there are a small number of galaxies that have star formation as the primary ionization source, but they have $3 < EW(H\alpha) < 6$ which suggests they are in the transition zone associated with the Green Valley. We have overplotted these galaxies in figures throughout the paper, but they are not incorporated into any statistical tests. As seen in Figure~\ref{fig:ew_gs}, these data typically lie towards low equivalent widths (as one would expect) but they generally fall along the fitted relation. This behavior will change somewhat in Section~\ref{sec:gs_ism_properties_corr}.

\begin{figure*}
    \centering
    \includegraphics[width=0.9\columnwidth]{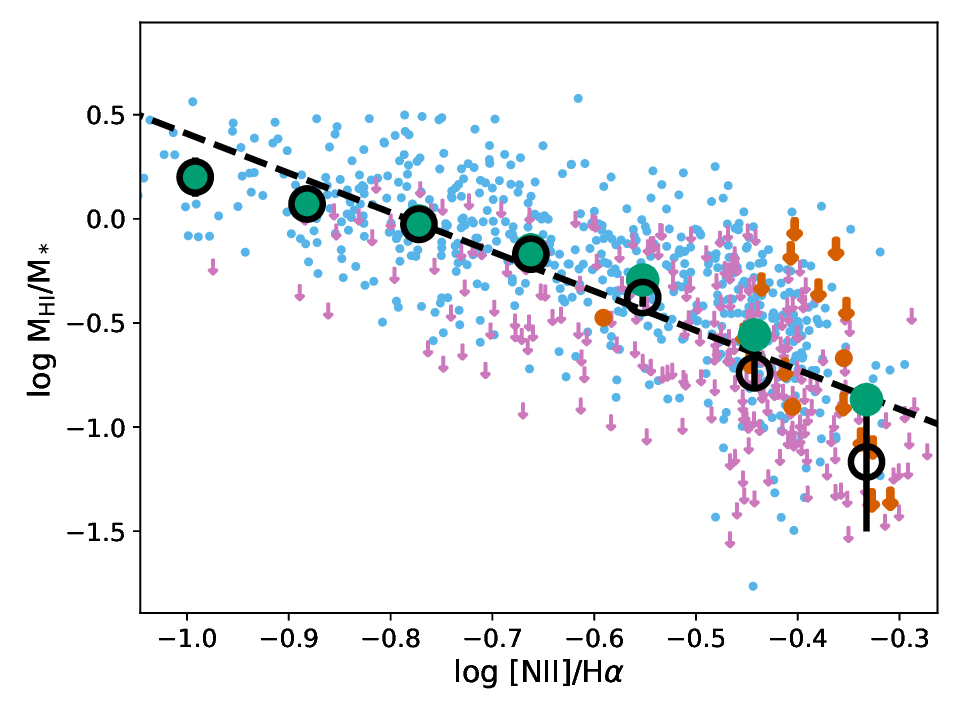}
    \includegraphics[width=.9\columnwidth]{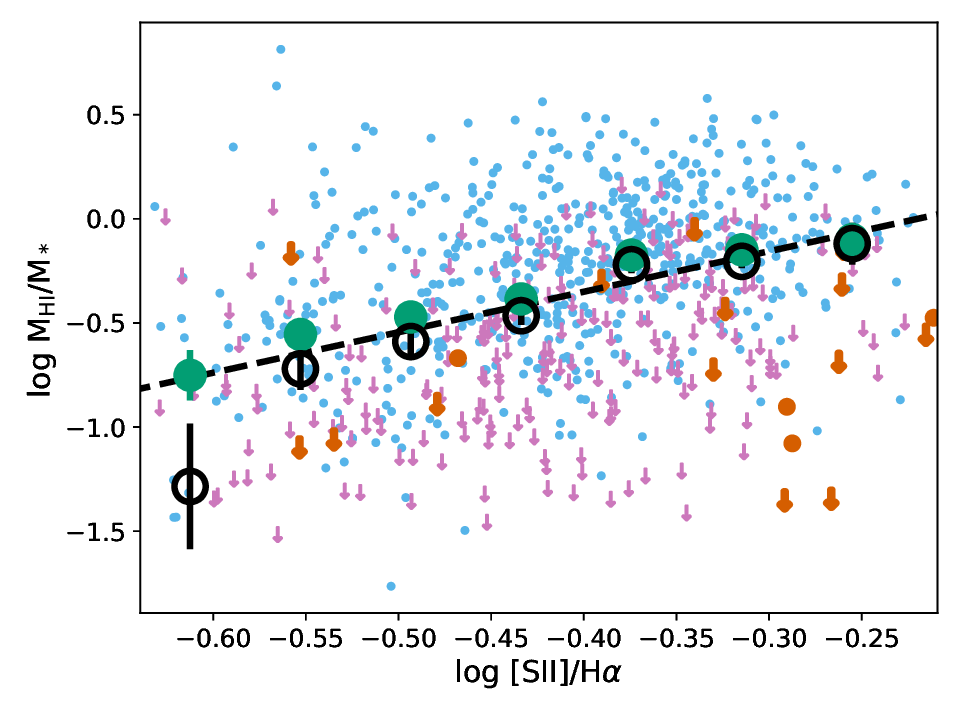}
    \includegraphics[width=.9\columnwidth]{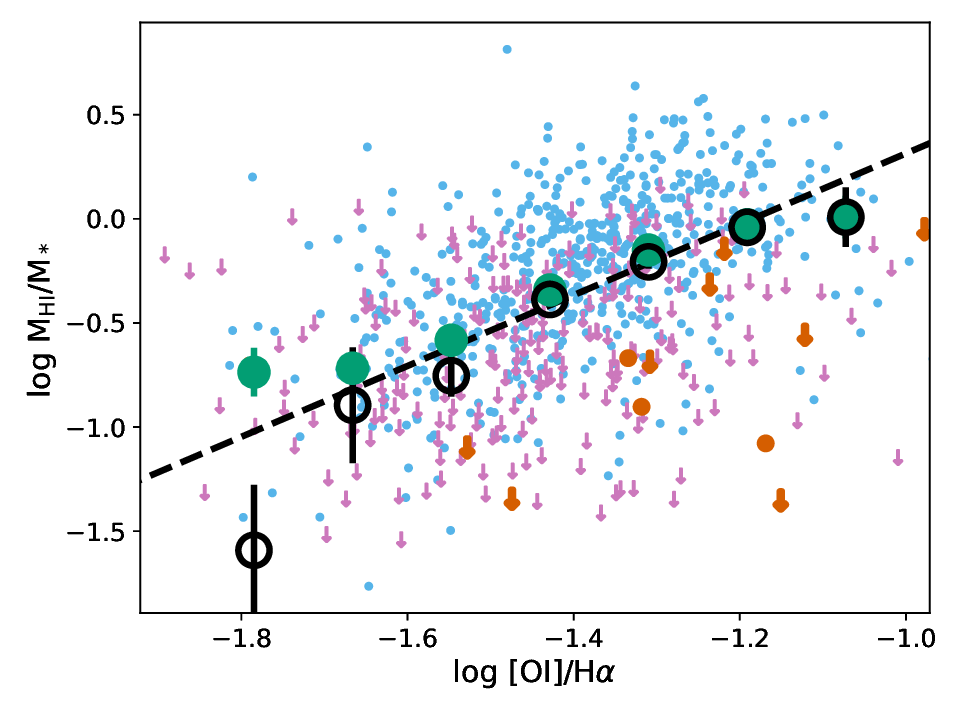}
    \includegraphics[width=.9\columnwidth]{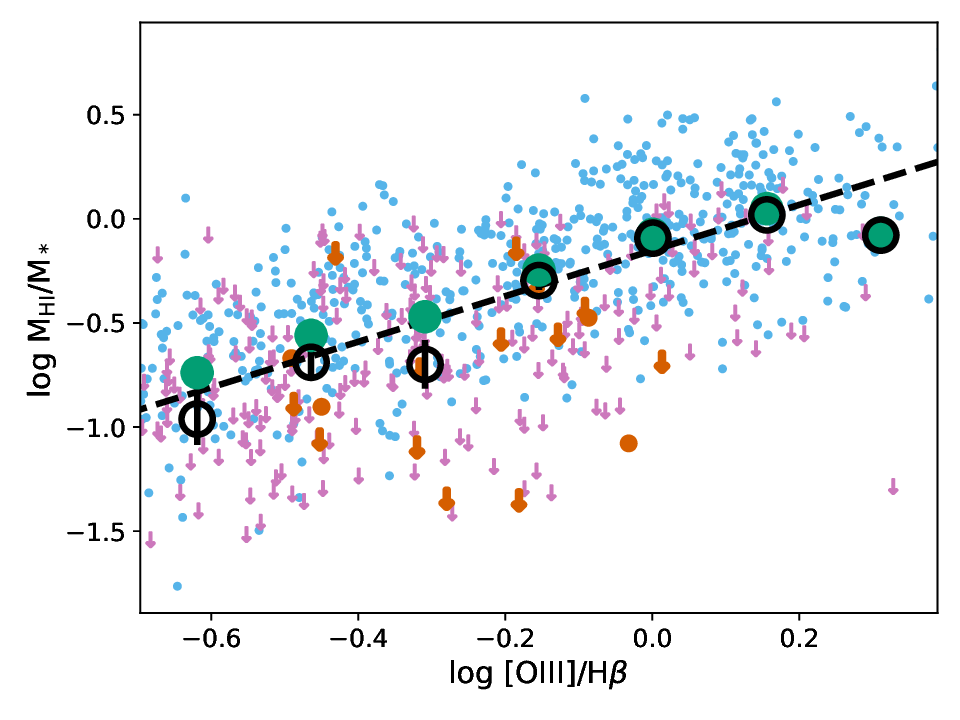}
    \includegraphics[width=.9\columnwidth]{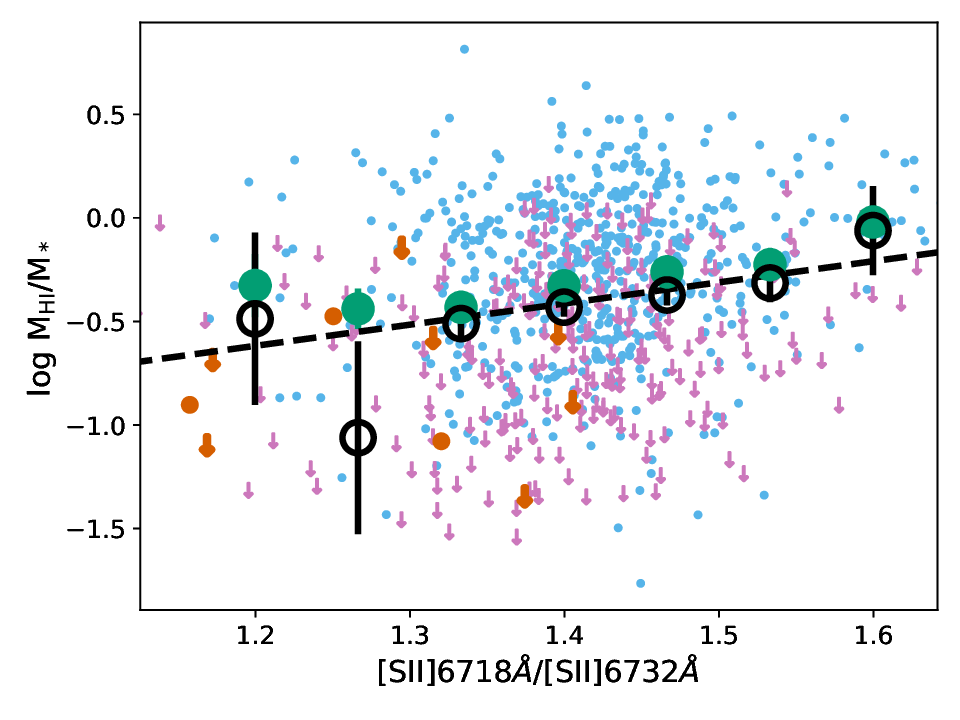}
    \includegraphics[width=.9\columnwidth]{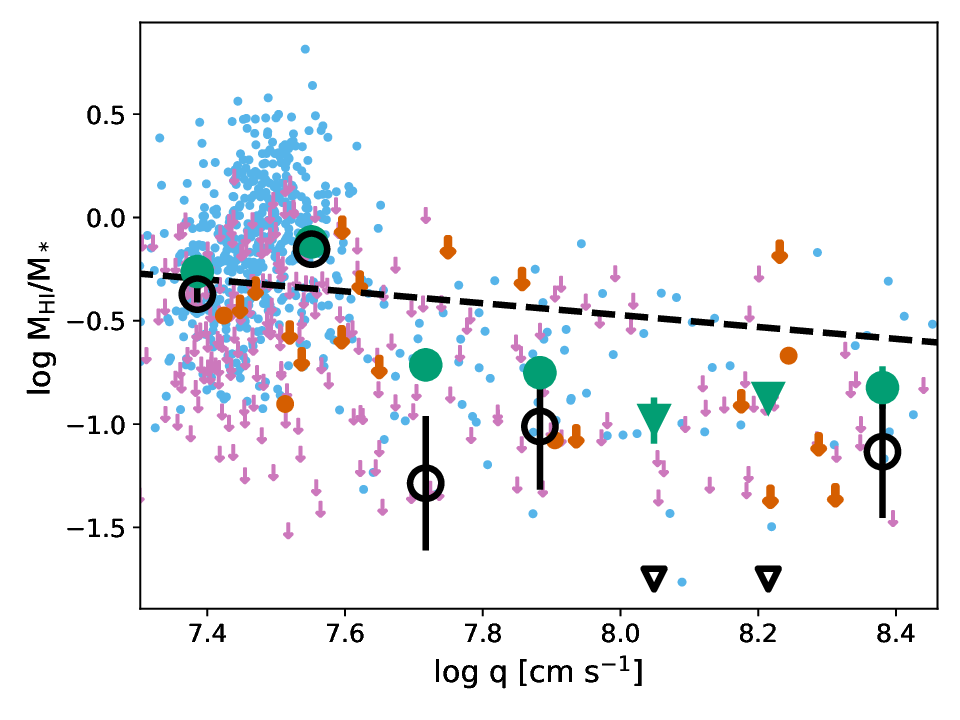}
    \includegraphics[width=.9\columnwidth]{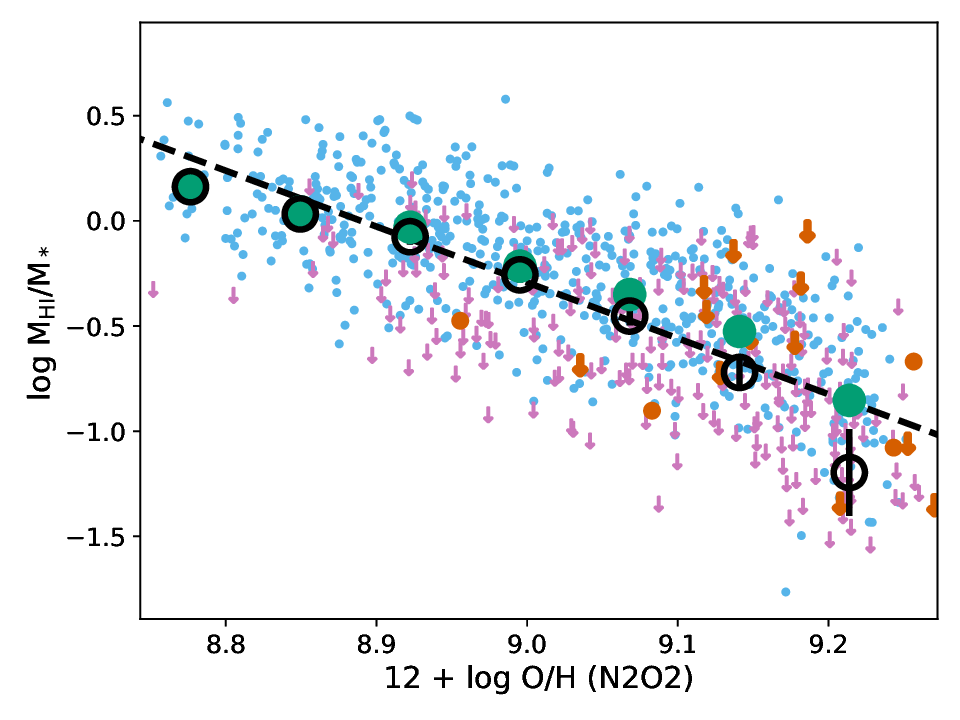}
    \includegraphics[width=.9\columnwidth]{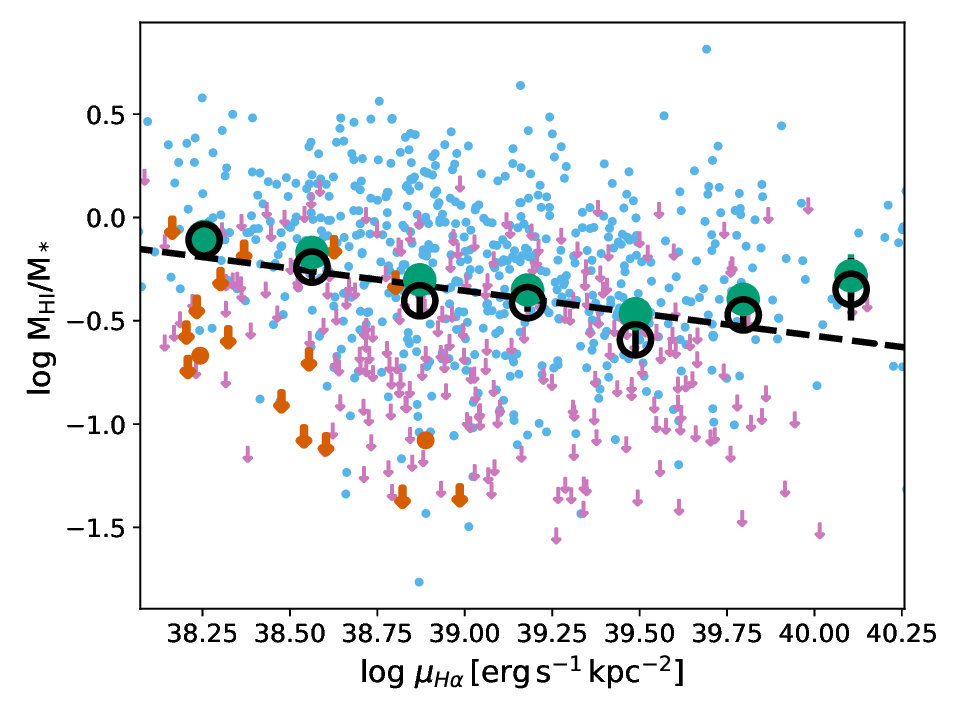}
    \caption{{\gs} vs. various ISM diagnostics based on optical emission lines ratios. Symbols are the same as in Figure~\ref{fig:ew_gs}.}
    \label{fig:gs_ratios}
\end{figure*}

\subsection{Correlations between $\text{M}_\text{H{\sc i}}/\text{M}_*$ and ISM diagnostics}
\label{sec:gs_ism_properties_corr}
Next we examine how {\gs} correlates with the other ISM diagnostics discussed in Section~\ref{sec:manga_properties}. The parameters describing each correlation are provided in Table~\ref{tab:scaling_tau} and all the relations are plotted in Figure~\ref{fig:gs_ratios}.

Again, statistically significant correlations are found with every single parameter, even though several correlation coefficients are quite small.  {\gs} has only a very mild dependence on [\ion{S}{ii}]/H$\alpha$, [\ion{S}{ii}]6718\AA/[\ion{S}{ii}]6732\AA, $q$, and $\mu_{H\alpha}$. Our analysis also highlights how the range of [\ion{S}{ii}]6718\AA/[\ion{S}{ii}]6732{\AA} is very narrow, implying roughly constant typical electron densities across all galaxies in our sample.  Similarly, our sample falls within a narrow range of $q$, except for a small subset of galaxies (typically at {\gs}$<1$) which skew towards higher values.

Much stronger correlations are found with [\ion{N}{ii}]/H$\alpha$, [\ion{O}{i}]/H$\alpha$, [\ion{O}{iii}]/H$\beta$, 12+log O/H.  The strong relationship between {\gs} and 12+log O/H is consistent with previous studies, and can be understood as relating to the correlation between {\gs} and stellar mass, the stellar mass-metallicity relation (MZR), and anticorrelation of the residuals in the MZR with {\hi} mass \citep{Lequeux79, Skillman96, Peeples08, Robertson12, Hughes13, Brown18}. Whether gas-phase metallicity is more fundamentally related to stellar mass, {\hi} mass, or {\gs} is beyond the scope of this work. 

Both [\ion{N}{ii}]/H$\alpha$ and [\ion{O}{iii}]/H$\beta$ are strongly correlated with metallicity, so [\ion{N}{ii}]/H$\alpha$, [\ion{O}{iii}]/H$\beta$, and 12+log(O/H) are all effectively illustrating the same {\gs}-metallicity relation. The metallicity dependence of [\ion{O}{i}]/H$\alpha$ is significantly weaker, and disappears below $12+\log{O/H}\sim9.05$ in our data. We rerun the correlation tests for [\ion{O}{i}]/H$\alpha$, limiting to $12+\log{O/H} < 9.05$, and find that the {\gs} vs. [\ion{O}{i}]/H$\alpha$ correlation persists at a statistically significant level (but with slightly lower correlation coefficient of 0.23 and p-value \num{9.6E-12}).

In contrast to Figure~\ref{fig:ew_gs}, the green valley points often skew below the best-fit line to star-forming galaxies. This behavior provides some idea of how the observed relations may vary as galaxies migrate off the star-forming sequence. However, examining the green valley population in further detail is beyond the scope of this work.

\subsection{Correlations with $\text{M}_\text{H{\sc i}}/\text{M}_*-$EW(O) residuals}
\label{sec:resid}
Given that the emissivity of optical emission lines are typically dependent on the local conditions of the ISM, we explore whether the scatter in the {\gs} vs. EW relations can be linked to variations in the {\it average} properties of the ISM. For this analysis, we focus our attention on the residuals, $\delta$\gs, of the strongest correlations using the three strong oxygen lines. We test for correlations between $\delta${\gs} and each of the ISM properties described in Section~\ref{sec:manga_properties}.

\begin{figure*}
    \centering
    \includegraphics[width=0.9\columnwidth]{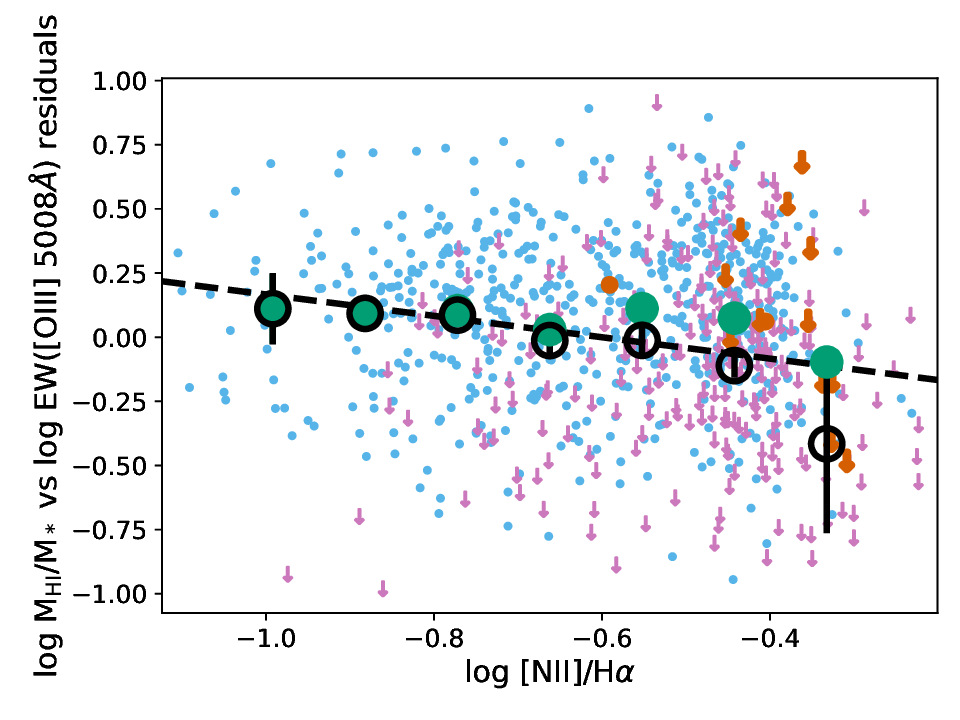}
    \includegraphics[width=0.9\columnwidth]{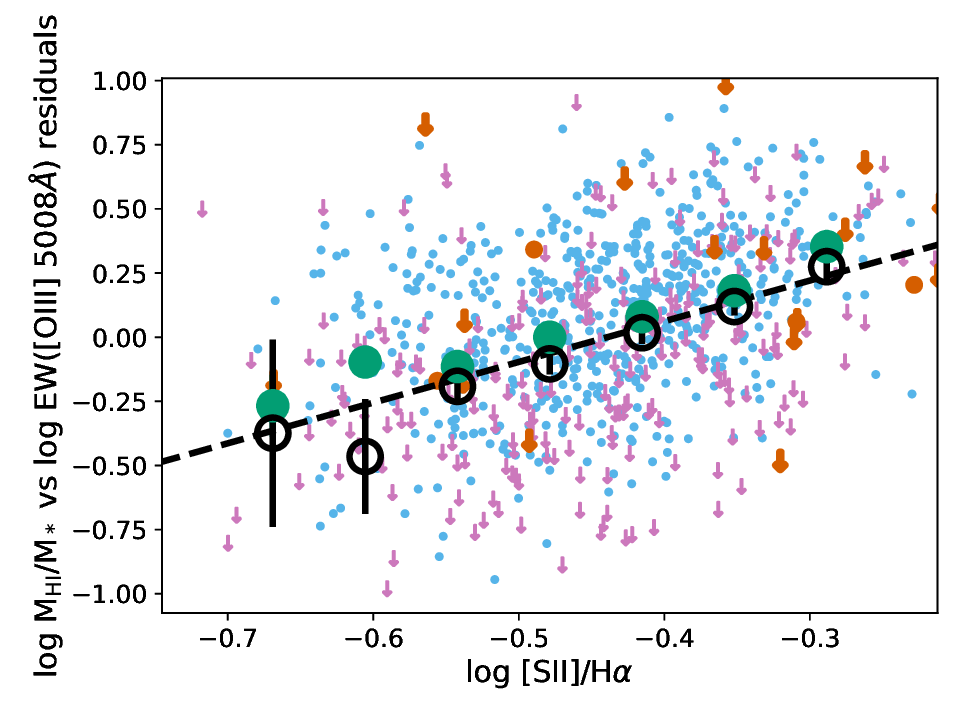}
    \includegraphics[width=0.9\columnwidth]{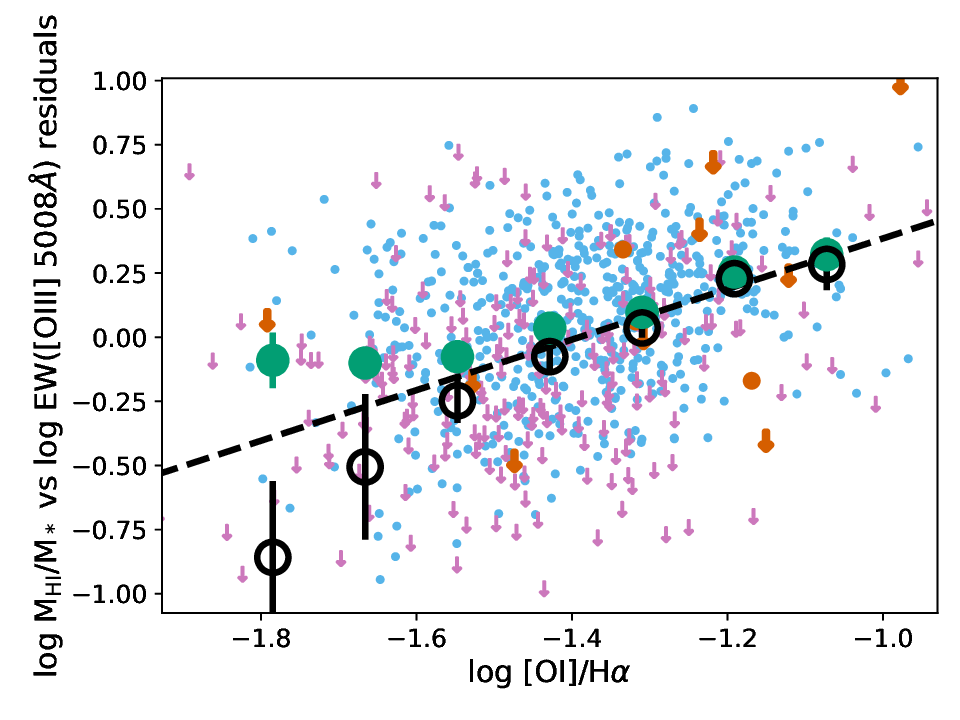}
    \includegraphics[width=0.9\columnwidth]{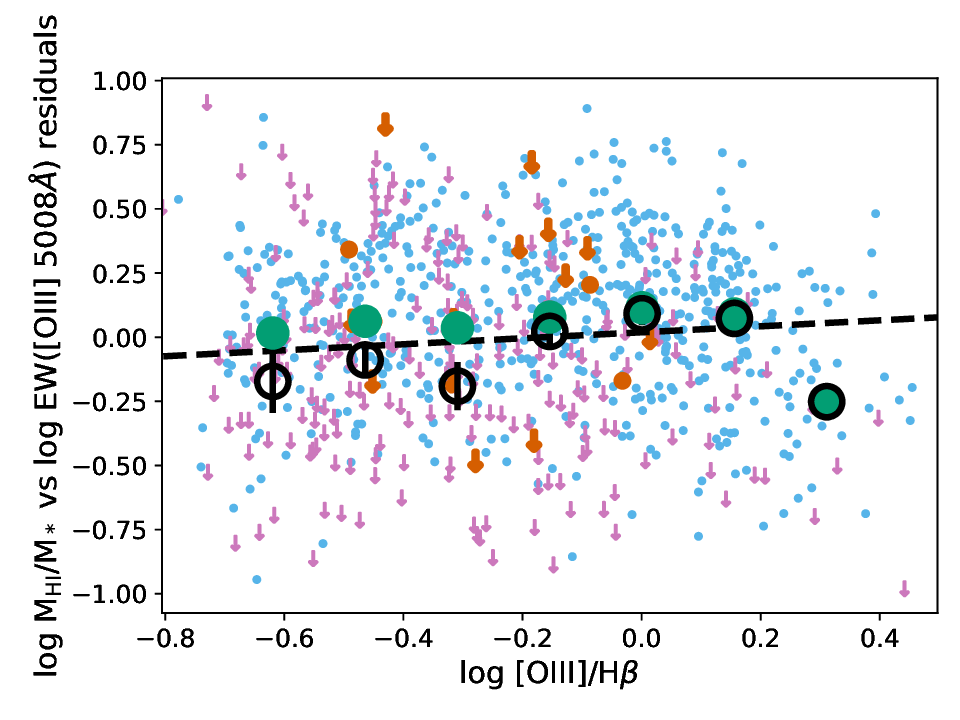}
    \includegraphics[width=0.9\columnwidth]{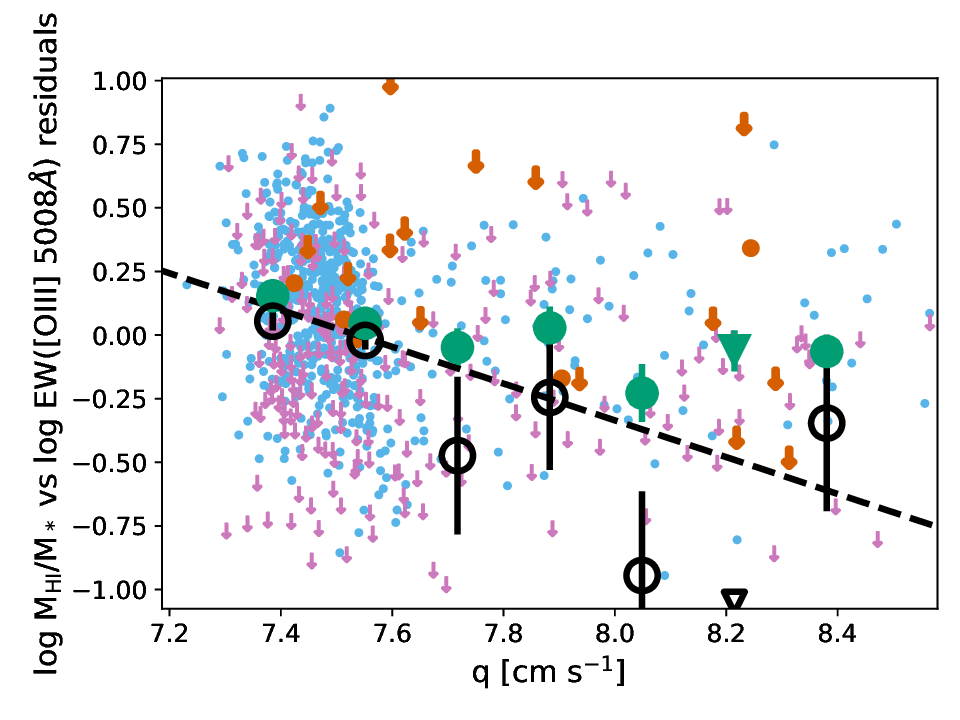}
    \includegraphics[width=0.9\columnwidth]{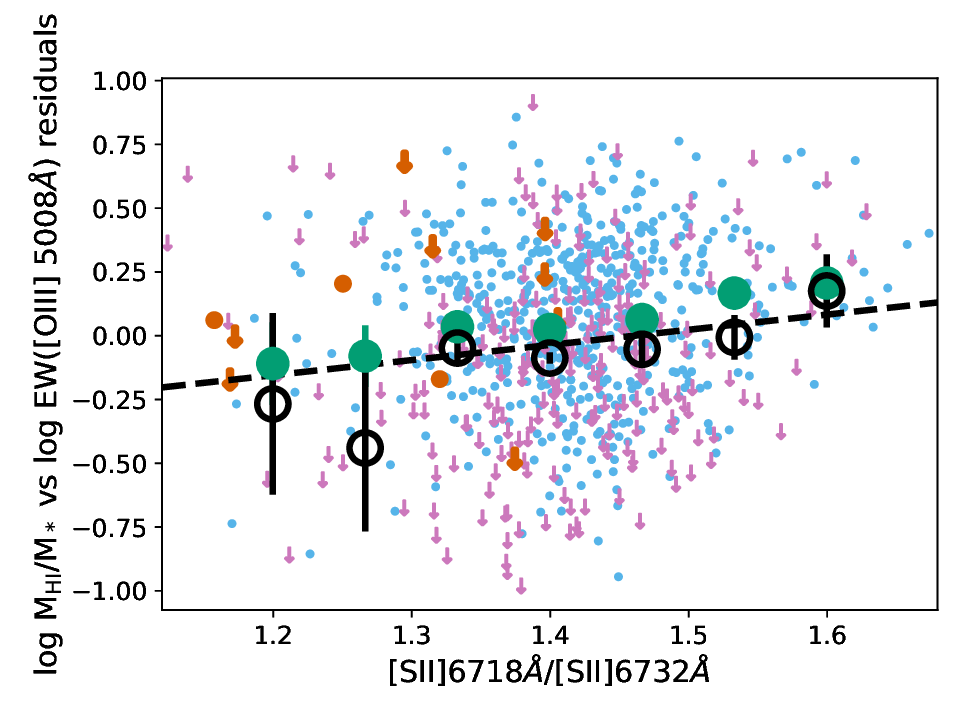}
    \includegraphics[width=0.9\columnwidth]{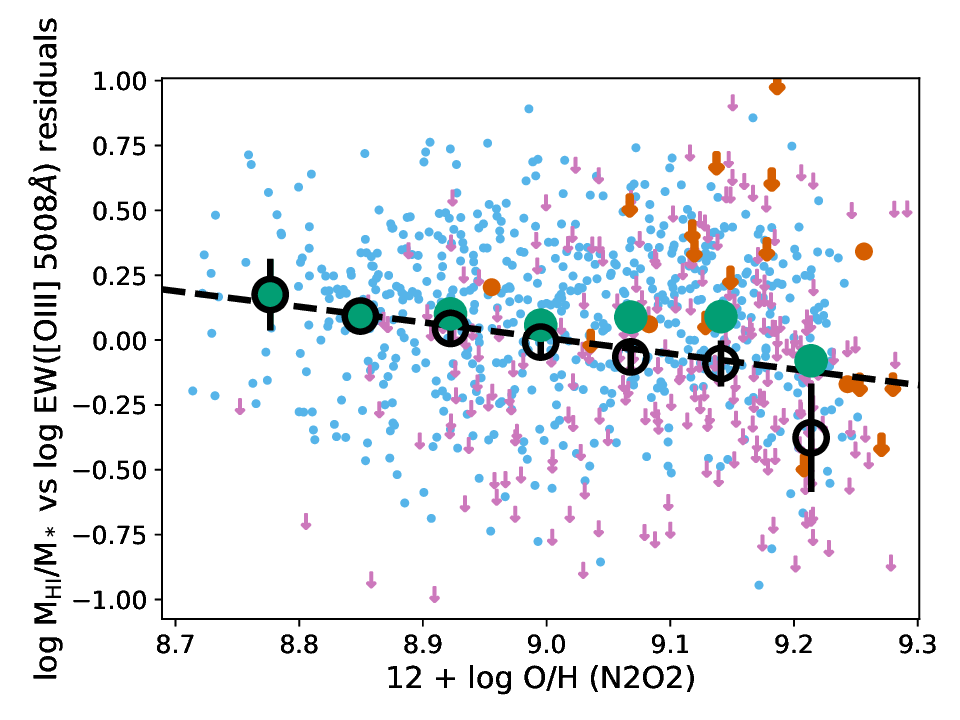}
    \includegraphics[width=0.9\columnwidth]{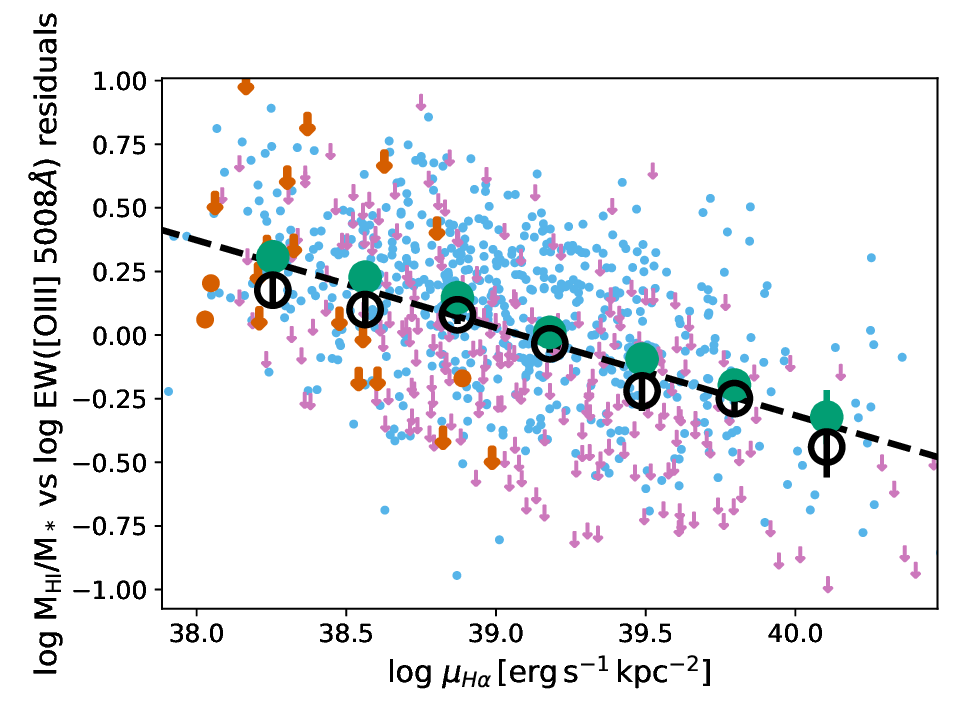}
    \caption{Correlations between the residuals of the {\gs}-EW([\ion{O}{iii}]) relation vs. various ISM diagnostics. Symbols are the same as in Figure~\ref{fig:ew_gs}.}
    \label{fig:gs_ew_residuals}
\end{figure*}

Table~\ref{tab:residuals_tau} provides Kendall's $\tau$ and p-value for correlations between $\delta${\gs} and the various ISM diagnostics, while Figure~\ref{fig:gs_ew_residuals} plots these residuals for the {\gs}-EW([\ion{O}{iii}]) relation. Correlation coefficients are generally low but statistically significant. The strongest correlations ($\tau\sim0.2-0.25$) are consistently found with $\mu_{H\alpha}$ and [\ion{O}{i}]/H$\alpha$, regardless of which {\gs}-EW(O) relation is considered. In some select cases, other ISM diagnostics show similarly strong correlations with $\delta${\gs} (e.g., the residuals when using the [\ion{O}{iii}] line are strongly correlated with {\siiha}) likely owing to the different conditions under which these lines originate. In contrast, correlations of $\delta${\gs} with $q$, [\ion{S}{ii}]6732\AA/[\ion{S}{ii}]6718{\AA}, and {\oiiihb} are consistently the weakest, with $|\tau|\lesssim0.15$. 

\begin{table*}
\caption{Correlation test $\tau$ and (p-value) for $\text{M}_\text{H{\sc i}}/\text{M}_*$ vs. EW residuals}
\label{tab:residuals_tau}
\begin{tabular}{lcccccc}   
\hline
& $[\ion{O}{ii}] 3729\AA$ & $[\ion{O}{iii}] 5008\AA$ & $[\ion{O}{i}] 6302\AA$ \\
\hline
$[\ion{N}{ii}]/H_{\alpha}$ &  -0.13(\num{1.02E-07}) &  -0.10(\num{1.23E-05}) &  -0.16(\num{2.69E-12})\\
$[\ion{S}{ii}]/H_{\alpha}$ &   0.13(\num{9.31E-09}) &   0.20($<$\num{2.22E-16}) &   0.17(\num{5.00E-13})\\
$[\ion{O}{i}]/H_{\alpha}$ &   0.21($<$\num{2.22E-16}) &   0.22($<$\num{2.22E-16}) &   0.19(\num{5.33E-15})\\
$[\ion{O}{iii}]/H_{\beta}$ &   0.08(\num{7.30E-04}) &   0.05(\num{4.36E-02}) &   0.14(\num{1.38E-08})\\
$[\ion{S}{ii}]6732\AA/[\ion{S}{ii}]6718\AA$ &   0.05(\num{2.71E-02}) &   0.06(\num{7.17E-03}) &   0.07(\num{2.06E-03})\\
$q$ &  -0.08(\num{5.12E-04}) &  -0.16(\num{3.76E-12}) &  -0.08(\num{5.48E-04})\\
$12+\log{O/H}$ (N2O2) &  -0.13(\num{2.84E-08}) &  -0.11(\num{2.77E-06}) &  -0.17(\num{5.58E-13})\\
$\mu_{H_{\alpha}}$ &  -0.22($<$\num{2.22E-16}) &  -0.23($<$\num{2.22E-16}) &  -0.24($<$\num{2.22E-16})\\
\hline
\end{tabular}
\end{table*}

\section{Discussion}
\label{sec:discussion}
In Section~\ref{sec:results}, we showed that out of all observed optical emission lines, {\gs} in star-forming galaxies correlates best with the EW of [\ion{O}{ii}], [\ion{O}{iii}], and [\ion{O}{i}], all to a similar degree.  Correlation strength with most other lines, most notably H$\alpha$, is significantly weaker. We also recover a strong anti-correlation between {\gs} and gas-phase metallicity, as well as a a positive correlation between {\gs} and [\ion{O}{i}]/H$\alpha$ (independent of metallicity). Furthermore, we found that the residuals of these scaling relations, $\delta${\gs}, correlate best with the mean $H\alpha$ surface brightness and  [\ion{O}{i}]/H$\alpha$.

In the following section we investigate and interpret these results in further detail.  In Section~\ref{sec:photogs_compare}, we compare our results to another analysis approach where {\hi} upper limits are replaced with estimates based on photometric parameters. In Section~\ref{sec:final_scaling}, we discuss whether optical spectroscopic data may be a useful means of indirectly estimating {\gs}, similar to how broadband colors are used. In Section~\ref{sec:radial_dep}, we explore how the aperture over which emission line properties are measured affects the resulting relations, specifically whether tighter scaling relations can be obtained with larger apertures. Lastly, in Section~\ref{sec:ism_properties}, we discuss what the observed trends may be telling us about the properties of the ISM in galaxies as a function of their gas-richness.  

\subsection{Comparison with photometric gas fractions}
\label{sec:photogs_compare}
As described in Section~\ref{sec:photogs}, another approach to incorporating upper limits into our analysis is to replace them with estimates of {\gs} based on existing scaling relations with already measured photometric properties.  Using the methods described in Sections~\ref{sec:photogs} and Appendix~\ref{app:photogs}, we replace {\hi} upper limits with new estimates and compare the derived correlation properties to those determined using the ATS estimator and our original upper limits.  

\begin{figure*}
    \centering
    \includegraphics[width=2\columnwidth]{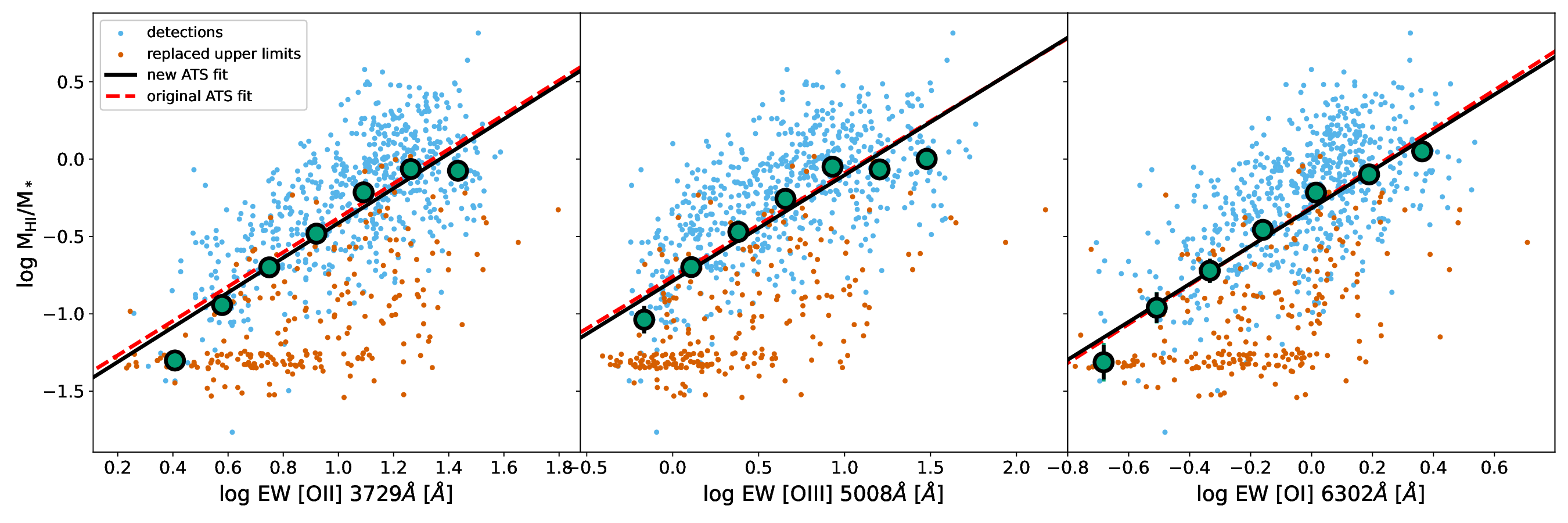}
    \caption{{\gs} vs. EW(O) correlations where upper limits have been replaced by new estimates of {\gs} using the photometric gas fraction technique (orange points). The black line shows the ATS fit to these data (where all points are treated as detections), while the red line shows the ATS fit to the data with the original upper limits. Green and black points are the same as in Figure~\ref{fig:ew_gs}, although this figure contains no upper limits so they are identical.}
    \label{fig:ewo_gs_photogs}
\end{figure*}

Figure~\ref{fig:ewo_gs_photogs} shows the {\gs}-EW(O) relations using the replaced upper limits. The black lines in Figure~\ref{fig:ewo_gs_photogs} represent the new linear fits, while the red lines represent the original linear fits from Section~\ref{sec:results} and Figure~\ref{fig:ew_gs}. Although using photometric gas fractions removes upper limits and the need for survival analysis, we still use the ATS estimator for the linear fits (with all data now treated as detections) to be as consistent as possible with our earlier analysis. The two fits are not drastically different. The values of $\tau$ when using the photometric-gas fraction approach are slightly higher (0.40$-$0.45) but generally consistent with the previous estimates, as are the measured upper and lower scatter ($\sigma_h$, $\sigma_l$), which are still asymmetric. The locus of points around {\gs}$\sim-1.3$ is artificial and represents the minimum value of {\gs} assumed by the photometric gas fraction method used here. If we rerun the ATS fit treating these values as upper limits, we obtain very similar results.  We avoid considering either the ATS or photometric gas fraction approach to upper limits to be  ``best", but our analysis illustrates that both approaches give consistent results.

\subsection[Multiparameter fits]{Using optical spectroscopy to predict $\text{M}_\text{H{\sevensize I}}/\text{M}_*$}
\label{sec:final_scaling}
A number of studies have identified tight correlations between {\gs} and broadband photometric properties, most notably color, but even tighter scaling relations have been formulated by incorporating additional parameters as well (e.g., surface brightness, axial ratio; see Section~\ref{sec:intro}). Here, we ask the question of whether similarly tight scaling relations can be formulated using spectroscopic information. One such attempt was made by \citet{Brinchmann13}, who was able to estimate {\it local} gas surface densities to within a factor of two using optical spectroscopy. However, they were unable to extend their approach to provide reliable estimates of integrated gas mass; their spectroscopic data was limited to the central 3$\arcsec$ of each galaxy, making the difference in spatial scales between the optical and 21cm data even more pronounced than in our study. The authors explored using aperture corrections, but found the systematic errors were too large to reliably estimate gas masses.

The tightest photometric gas fraction estimators have scatters of $\sim0.3$, or a factor of two \citep[e.g.][]{Catinella13,Eckert15}, although estimators calibrated on a sample of purely star forming galaxies without AGN (like in this work) may have even smaller scatter. A clear disadvantage of the scaling relations in this work is the asymmetric scatter. Although the upper scatter approaches 0.3 dex in some cases, the corresponding lower scatter typically $>0.4$ dex.

\begin{figure*}
    \centering
    \includegraphics[width=2\columnwidth]{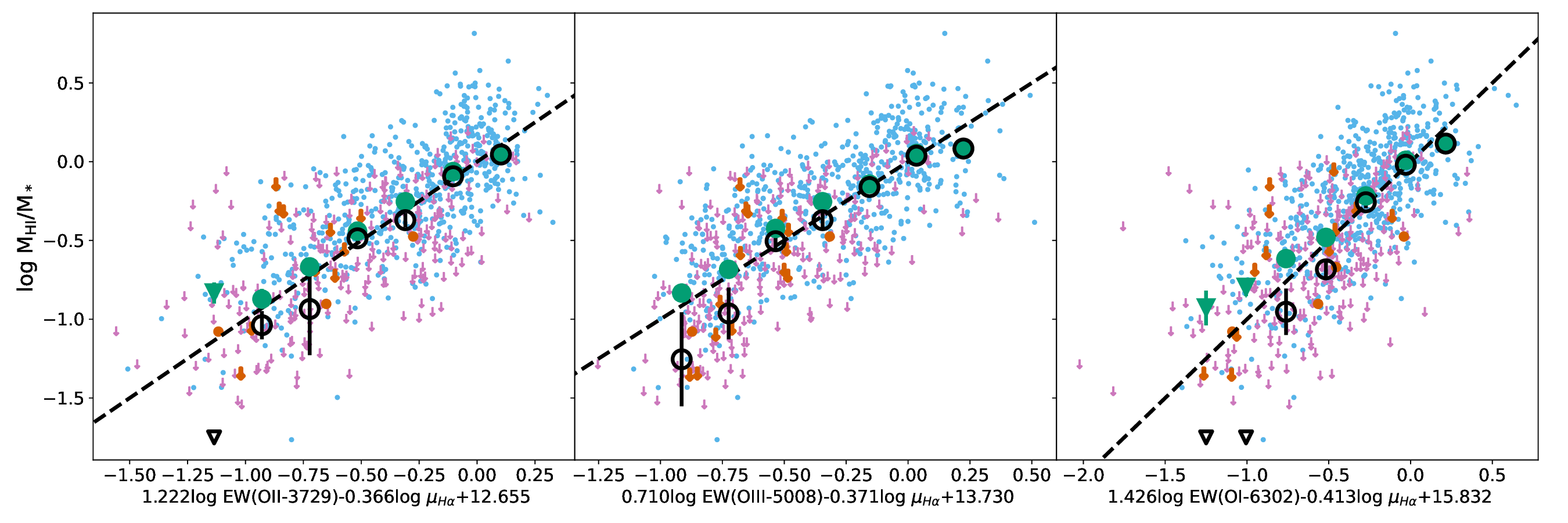}
    \caption{Final {\gs} scaling relations combining EW(O) with $\mu_{H\alpha}$. Symbols are the same as in Figure~\ref{fig:ew_gs}.}
    \label{fig:final_corr}
\end{figure*}

By combining the strongest correlations from Section~\ref{sec:gs_ew} and Section~\ref{sec:resid}, we explore whether we can create stronger, more symmetric scaling relations between {\gs} and optical emission line properties. For this purpose we combine EW(O) and $\mu_{H\alpha}$, which consistently correlates most strongly with the residuals of the {\gs}-EW(O) relations. The final relations, shown in Figure~\ref{fig:final_corr}, are:
\begin{multline}
\label{eq:final1}
    \log{M_{\ion{H}{i}}/M_*} = (1.222\pm0.079)\log{EW([\ion{O}{ii}]3729)} \\
    + (-0.366\pm0.034)\log{\mu_{H\alpha}}
    +(12.655\pm1.320),
\end{multline}
\begin{multline}
\label{eq:final2}
\log{M_{\ion{H}{i}}/M_*}= (0.710\pm0.035)\log{EW([\ion{O}{iii}]5008)} \\
+(-0.371\pm0.031)\log{\mu_{H\alpha}} +(13.730\pm1.194),
\end{multline}
\begin{multline}
\label{eq:final3}
\log{M_{\ion{H}{i}}/M_*}= (1.426\pm0.080)\log{EW([\ion{O}{i}]6302)} \\
+(-0.413\pm0.039)\log{\mu_{H\alpha}} +(15.832\pm1.521).
\end{multline}
The upper/lower scatter ($\sigma_h$/$\sigma_l$) are 0.30/0.40, 0.31/0.40, and 0.32/0.41 for equations~\ref{eq:final1}, \ref{eq:final2}, and \ref{eq:final3}, respectively. The correlation coefficients are 0.43$-$0.46.

The above relations still suffer from asymmetric scatter, although the asymmetry has been substantially lessened by including a third parameter. Although the {\gs}-{\niiiha} and {\gs}-({$12+\log{O/H}$}) relations have similar values of $\tau$ and comparable scatter using only one parameter, the relations using EW(O) and $\mu_{H\alpha}$ have some notable advantages: they require no extinction corrections, the lines are less likely to be blended, and they are not dependent on specific photoionization models. Nonetheless, for the purposes of using optical properties to predict the {\hi} content of galaxies, existing relations in the literature using broadband color are likely more reliable at this time. Improvements may be possible, however.  For example, deeper {\hi} data may allow us to better understand the scatter and whether it varies across parameter space, and using optical emission line measurements over larger areas (or applying aperture corrections) is likely to tighten these relations (see Section~\ref{sec:radial_dep}). 

\subsection{Dependence on radial spectroscopic coverage}
\label{sec:radial_dep}

Throughout our analysis, we have been using \ion{H}{i} measurements from single-dish observations, where the beams (FWHM=3.5\arcmin--9.1\arcmin) are significantly larger than the  individual galaxies and likely contain {\it all} of their \ion{H}{i} emission.  In contrast, our emission line measurements are conducted out to $R_e$. Although the MaNGA IFU coverage of the Primary+ sample nominally extends to 1.5$R_e$, there are slight variations in coverage, so $R_e$ was chosen as a scale over which measurements can be conducted consistently for all galaxies.  However, \ion{H}{i} disks are frequently observed to extend well beyond this radius \citep{Broeils97}. It is possible that the difference in spatial scales being probed is contributing to at least some of the scatter in the relationships presented in Section~\ref{sec:results}, unless all \ion{H}{i} disks are self-similar such that they always contains the same fraction of \ion{H}{i} inside $R_e$, but this is unlikely. Although \citet{Broeils94} and \citet{Swaters02} do find similarities in radial \ion{H}{i} distributions at fixed morphology, there is significant scatter around the mean distributions, with an additional dependence on environment also possible \citep{Cayatte94}.

To explore this issue further, we compare the {\gs} vs. EW([\ion{O}{i}]) correlation using emission line measurements conducted out to $R_e/2$, $R_e$, and $2R_e$.  The line measurements out to $2R_e$ are not done uniformly; we simply take any data in the IFU which lies within $2R_e$. Typically, the data at larger radius falls along the minor axis and is visible within the IFU due to the a galaxy's inclination. Although these measurements are not uniform, if a larger aperture improves the {\gs} scaling relations, we should still see some sign of it, although the overall improvement may be underestimated. [\ion{O}{i}] is used here as an example; the results of this analysis also apply to the other lines.

Table~\ref{tab:fit_radius} gives the linear fit parameters to the \gs-EW([\ion{O}{i}]) and the $\delta${\gs}-$\mu_{H\alpha}$ relations, as well as the final upper and lower scatter in the combined scaling relation. Increasing the emission line measurement aperture does appear to mildly strengthen the \gs-EW([\ion{O}{i}]) correlation strength, although it levels off once an aperture of $R_e$ is reached. The derived slope and intercept also change significantly. The strength of the correlation between $\delta${\gs} and $\mu_{H\alpha}$ is essentially independent of IFU radial coverage, and while the derived slopes/intercepts change, they are in agreement within their error bars. Notably, increasing IFU radial coverage decreases both the size of the scatter and the asymmetry in the relations shown in Table~\ref{tab:fit_radius}, implying the different spatial scales over which 21cm and optical emission is measured is contributing to the scatter. 

\begin{table}
    \centering
    \caption{Fit parameters for different IFU radial coverage}
    \begin{tabular}{lccc}
    \hline
     & $R_e/2$ & $R_e$ & $2R_e$ \\
    \hline
     \multicolumn{4}{c}{{\gs} vs. EW([\ion{O}{i}])6302\AA} \\
    \hline
    $\tau$ &   0.29 &   0.36 &   0.33\\
slope &   0.90$\pm$  0.05 &   1.26$\pm$  0.07 &   1.46$\pm$  0.12\\
intercept &  -0.23$\pm$  0.02 &  -0.31$\pm$  0.02 &  -0.39$\pm$  0.03\\
$\sigma_h$ &   0.39 &   0.37 &   0.32\\
$\sigma_l$ &   0.57 &   0.49 &   0.43\\
    \hline
    \multicolumn{4}{c}{$\delta${\gs} vs. $\mu_{H\alpha}$} \\
    \hline
    $\tau$ &  -0.25 &  -0.24 &  -0.25\\
slope &  -0.38$\pm$  0.03 &  -0.36$\pm$  0.03 &  -0.35$\pm$  0.03\\
intercept &  14.99$\pm$  1.13 &  14.27$\pm$  1.28 &  13.42$\pm$  1.10\\
$\sigma_h$ &   0.33 &   0.31 &   0.30\\
$\sigma_l$ &   0.52 &   0.43 &   0.38\\
     \hline
     \multicolumn{4}{c}{{\gs} vs. EW([\ion{O}{i}]6302\AA)$+\mu_{H\alpha}$} \\
     \hline
     $\sigma_h$ &   0.35 &   0.32 &   0.30\\
$\sigma_l$ &   0.49 &   0.41 &   0.36\\
    \end{tabular}
    \label{tab:fit_radius}
\end{table}

As a consistency check, we conduct the same analysis as above but using the MaNGA Secondary sample (with radial coverage out to $\sim2.5R_e$) where spectroscopic measurements to $2R_e$ can include all data along the major axis, but with the caveat that this subsample suffers from a smaller sample size and a larger rate of non-detections due to its higher redshift distribution.  We find similar qualitative results as described above, although all fit parameters suffer from larger errors.

From this analysis, we conclude that the correlations identified in Section~\ref{sec:results} persist regardless of the area over which the emission lines are measured, but measuring emission line properties over larger areas does appear to increase the correlation strength and decrease the scatter. The improvement when using an aperture of $2R_e$ may actually be underestimated due to our non-uniform measurements out to this radius. Given the tightening of the final relation when larger apertures are used, it may be valuable to explore in future work whether using even larger-area IFU bundles or applying careful aperture corrections can yield even stronger relations. 

\subsection{Gas Content and average ISM properties}
\label{sec:ism_properties}

Our results provide a means of understanding the typical ISM conditions in star-forming galaxies as a function of their gas-richness.  We again stress that we are discussing {\it average} ISM properties, and the local conditions within galaxies almost certainly vary significantly around the average behavior.  Furthermore, as was discussed in Section~\ref{sec:radial_dep}, the optical emission and {\hi} emission are being measured over significantly different areas. While the optical emission is measured within $r$-band effective radius ($R_e$), \citet{Broeils97} find that the typical {\hi} effective radius roughly equals the $B$-band 25 mag arcsec$^{-2}$ isophotal radius, and thus the majority of {\hi} emission comes from beyond $R_e$. Thus, we are not measuring the conditions of all the {\hi} in our sample. Rather, we are examining the ISM conditions in the star-forming disks of galaxies, and assessing how these relate to {\hi}-richness.

In agreement with previous work, we find a clear anticorrelation between gas content and metallicity \citep{Lequeux79, Skillman96, Peeples08, Robertson12, Hughes13, Brown18}. This result is implied both through our estimate of metallicity using the N2O2 predictor as well as [\ion{N}{ii}]/H$\alpha$ and [\ion{O}{iii}]/H$\beta$ which are very sensitive to metallicity. The gas-phase metallicity shows the strongest variation with {\gs} of all properties estimated from emission lines.  In contrast, [\ion{S}{ii}]6718\AA/[\ion{S}{ii}]6732{\AA} varies only slightly, with most of the data lying around $\sim$1.4-1.5.  [\ion{S}{ii}]6718\AA/[\ion{S}{ii}]6732{\AA} is  minimally dependent on electron density at these values \citep{Osterbrock06}, but most galaxies are consistent with having average electron densities $\lesssim100\,{\rm cm^{-3}}$. Systematic variations in electron densities may have impacted other estimates of ISM properties which make assumptions about electron density, but thankfully this is not the case. Ionization parameter, $q$, similarly tends to cluster around a narrow range of values ($\sim10^{7.5}\,{\rm cm\,s^{-1}}$).  The few galaxies which have $q$ at higher levels notably tend to only exist at \gs$< 1$.  

We also find a correlation between {\gs} and  [\ion{O}{i}]/H$\alpha$, even after removing the metallicity dependence of [\ion{O}{i}]/H$\alpha$.   [\ion{O}{i}]/H$\alpha$ is sensitive to ionization fraction and gas temperature \citep{Reynolds98}. Its enhancement in gas-rich galaxies may indicate an on-average harder ionizing spectrum incident on the emitting gas, or more excitation from shocks \citep{Dopita00,Kewley06}. Increased shock excitation would be consistent with the higher turbulent velocities expected in lower mass galaxies \citep{Dalcanton04}.  This trend may also 
be consistent with a larger contribution of optical line emission from DIG, especially given the additional (albeit very weak) correlation between {\gs} and $\mu_{H\alpha}$.

\subsubsection{Why does $\text{M}_\text{HI}/\text{M}_*$ correlate more strongly with integral color than H$\alpha$ emission?}
\label{sec:gs_halpha_why}

{\gs} is only weakly correlated with $EW(H\alpha)$, a finding that is consistent with previous work by \citet{Jaskot15}. This result implies a relatively loose connection between the instantaneous specific star formation rate (H$\alpha$ traces star formation on $\sim$5 Myr timescales; \citealt{Leroy12}, \citealt{Kennicutt12}) and the size of the {\hi} reservoir. The weak correlation between {\gs} and EW(H$\alpha$) is in stark contrast to the significantly tighter relationship between {\gs} and color\footnote{``Color" in this discussion refers to any of the colors using a combination of near-ultraviolet, optical, and near-infrared passbands that have been shown to correlate well with {\gs}, e.g., ${\rm NUV}-r$, $u-J$, $u-r$, $g-r$ \citep{Kannappan13,Catinella13,Jaskot15,Eckert15}}, even though both color and EW(H$\alpha$) can be thought of as tracers of sSFR. We consider a number of possible explanations for the large scatter in the {\gs}-EW(H$\alpha$) relation, especially in contrast to the {\gs}-color relation: (1) scatter in the relationship between EW(H$\alpha$) and sSFR,  (2) $EW(H\alpha)$ and color measured over different apertures, (3) dust extinction artificially tightening the {\gs}-color relation, and (4) the different timescales over which $EW(H\alpha)$ and color trace sSFR.

We have been assuming that EW(H$\alpha$) has a 1:1 relationship with sSFR.  EW(H$\alpha$) is the H$\alpha$ line flux normalized by the stellar continuum flux density, while sSFR is star formation rate normalized by stellar mass. The mapping between stellar continuum flux density and stellar mass may depend on the details of the stellar population (age, metallicity, IMF). Similarly, the H$\alpha$ luminosity to SFR conversion makes a number of assumptions about the stellar IMF, gas metallicity, gas temperature, and gas density \citep{Kennicutt98b}, and if any of these assumptions are not true, then they might contribute to a more scattered {\gs}-EW(H$\alpha$) relation. A full analysis of the possible error in the H$\alpha$ luminosity to SFR conversion is beyond the scope of this work, but we do analyze the sSFR-{\gs} correlation using extinction-corrected H$\alpha$-based SFR from \texttt{Pipe3D} and stellar masses from the NSA, and we find the correlation coefficient is almost exactly the same as when using EW(H$\alpha$).

\begin{figure*}
    \centering
    \includegraphics[width=2\columnwidth]{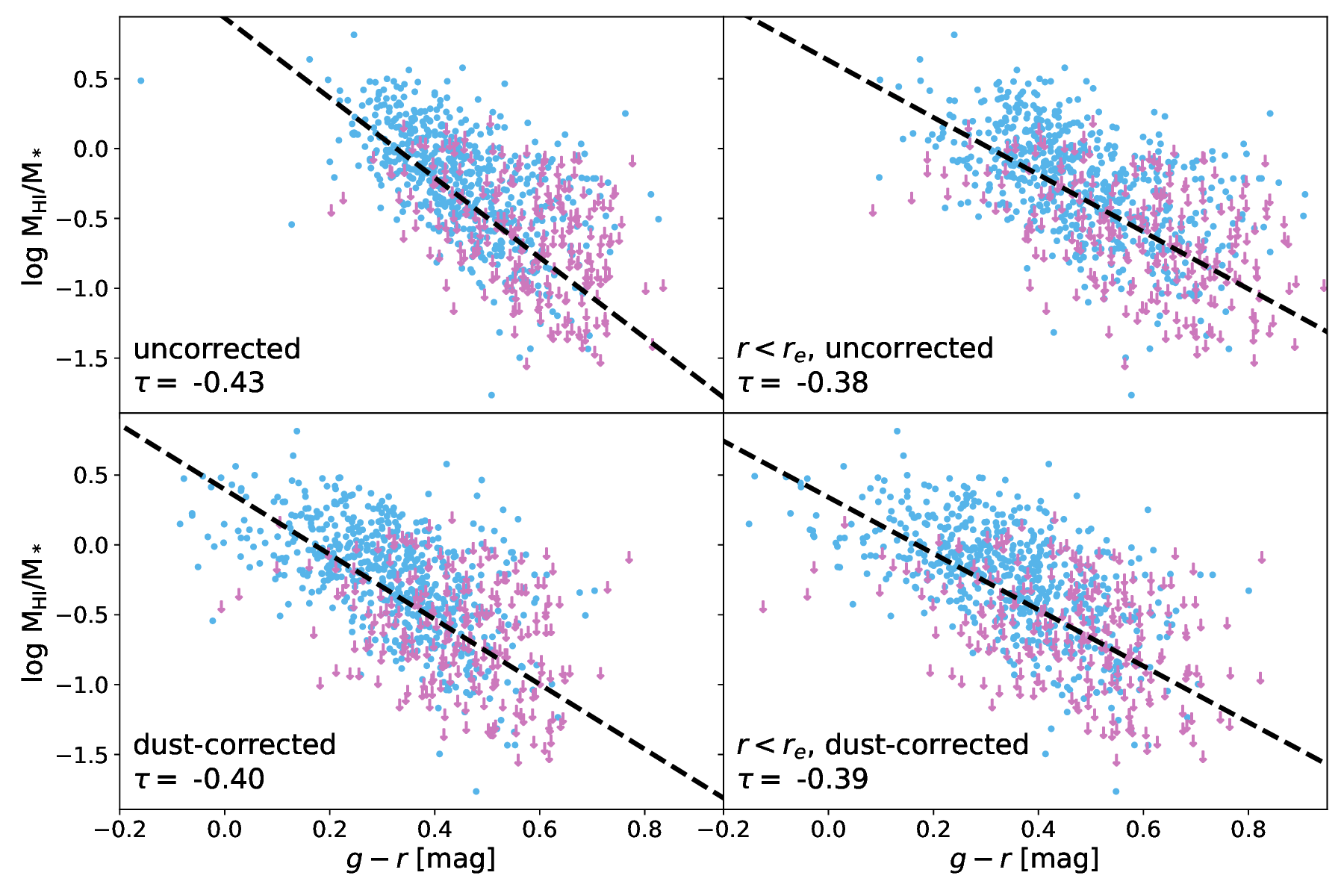}
    \caption{{\gs} vs. $g-r$ before and after correcting for internal extinction and/or limiting the color measurements to within $r_e$. Neither correction, nor the combination of the two, lead to a relationship as weak as the {\gs}-EW(H$\alpha$) relation.}
    \label{fig:gs_color}
\end{figure*}

The limited radial coverage of our spectroscopic measurements may drive at least some scatter in the {\gs}-EW(H$\alpha$) relation. Our analysis in Section~\ref{sec:radial_dep} implies that increasing the area over which optical spectroscopic properties are measured should tighten relations with {\gs}. However, increasing the radial coverage of the spectroscopic measurements is unlikely to account for all the scatter; \citet{Jaskot15} find a similarly weak {\gs}-EW(H$\alpha$) relation using narrow band H$\alpha$ imaging where emission can be measured globally. Nonetheless, we examine the change in correlation strength when it is measured out to 1.5$R_e$, i.e., the radial extent of the MaNGA Primary+ IFU bundles. $\tau$ increases only slightly from 0.13 to 0.16, but we cannot obtain a truly global EW(H$\alpha$) measurement due to the limited IFU sizes.  Therefore, we reverse the problem, and examine whether the {\gs}-color relation becomes a much weaker correlation if the colors are measured within $R_e$, using the $g-r$ color derived from images from the NSA (Figure~\ref{fig:gs_color}). The original relationship using global $g-r$ has $\tau=-0.43$ while the version with $g-r$ measured within $R_e$ has $\tau=0.38$. We conclude that the limited radial coverage of our spectroscopic measurements can explain some but not all of the weakness of the {\gs}-EW(H$\alpha$) relationship.

\citet{Jaskot15} argue that dust extinction artificially tightens the {\gs}-color relationship, and when applying internal extinction corrections, they obtain a correlation with {\gs} that has scatter similar to that of the {\gs}-EW(H$\alpha)$ relation. However, we argue that \mbox{\citet{Jaskot15}} may have {\it over-corrected} the broadband colors for internal dust extinction.  Specifically, they estimate internal extinction from the WISE mid-infrared (MIR) dust emission, which will be weighted towards the densest, and thus more extincted, regions of the ISM. However, broadband colors can reflect long-term growth rates of galaxies on timescales of up to Gyrs \citep{Kannappan13}, so the stellar light is coming from stars which are very likely no longer embedded in their birth clouds and not as extincted as the emission coming from more active star forming regions.  

An alternative approach to determine internal extinction corrections is to conduct a statistical analysis of galaxy colors as a function of axial ratio (a proxy for inclination) and absolute magnitude (or stellar mass), under the assumption that the reddening will be more pronounced for edge-on galaxies and more massive galaxies with higher metallicities \citep[][]{HernandezToledo08, Masters10}. We apply the statistical internal extinction corrections of \citet{HernandezToledo08} to the $g-r$ colors of our sample. The correlation strength is only marginally weakened with the addition of dust corrections, with $\tau$ changing from -0.43 to -0.40, and the weaker correlation is still much stronger than the {\gs}-EW(H$\alpha$) relation (Figure~\ref{fig:gs_color}). As a final test, we assess the strength of the {\gs}-EW(H$\alpha$) relation with both the dust corrections and colors measured out to $R_e$, but find the combination still does not significantly affect the correlation strength (Figure~\ref{fig:gs_color}).

Having largely ruled out more mundane explanations of the weak relationship between {\gs} and EW(H$\alpha$), we now explore whether there is a more physical explanation for the loose association between H$\alpha$ emission and {\hi} content. A weak relationship between these quantities may not actually be surprising when considering previous work on the star formation law ($\Sigma_{\rm SFR}$ vs. $\Sigma_{\rm gas}$, where $\Sigma$ refers to surface density) in nearby galaxies, where the results clearly show active star formation rate is most closely correlated with the amount of {\it molecular} hydrogen (H$_2$), not the amount of {\hi} \citep{Bigiel08}.  This result holds even when the two phases of hydrogen are compared at fixed surface density \citep{Schruba11}. If the global H$_2$/{\hi} were constant, or smoothly varying as a function of {\gs}, then the {\gs}-EW(H$\alpha$) relation might appear tighter. However, \citet{Catinella18} and \citet{Calette18} illustrate that H$_2$/{\hi} scaling relations have very large scatter, and H$_2$/{\hi} are more closely correlated with properties like stellar surface mass density and bulge strength. Other studies have also highlighted how H$_2$/{\hi} can change significantly due to interactions \citep{Kenney89, Braine93, Lisenfeld11, Stark13} and bar inflows \citep{Sakamoto99, Sheth05}. On small scales within galaxies, molecular cloud formation is likely dependent on a number of local conditions, including the overall gas density, details of the ionizing radiation field, and gas-phase metallicity \citep{Krumholz09}.  In summary, although {\hi} ``fuels" star formation by providing a gas reservoir from which molecular clouds can form, the detailed and complicated physics involved in converting \ion{H}{i} into H$_2$ on short timescales lead to a very messy {\gs}-EW(H$\alpha$) relation.

Instead, we suggest the difference in timescales probed by EW(H$\alpha$) and various colors explains their different correlation strengths with {\gs}.  EW(H$\alpha$), being sensitive on 5 Myr timescales, most strongly correlates with the properties of the gas where those stars are actively forming, namely the overall mass density of H$_2$. Meanwhile, global colors typically trace star formation over much longer timescales. For example, \citet{Kannappan13} show that U-NIR color (referring to a near ultraviolet minus a near infrared color, e.g., $u-J$) is a good proxy for the relative increase in stellar mass over the last Gyr\footnote{The growth rate used in \citet{Kannappan13} is defined as new stars formed in the last Gyr divided by the pre-existing stellar mass, so it is not exactly sSFR like used throughout our work where the denominator is total stellar mass including recently formed stars.}. Similarly, ${\rm NUV}-r$ should trace sSFR on timescales of up to $\sim200$ Myrs \citep{Kennicutt12}. Therefore, although stars do not form in {\hi}, it is the {\hi} reservoir (and its likely regular replenishment) which {\it sustains} star formation by continually providing the material out of which molecular clouds and stars form, thus driving the {\gs}-color relationship.
 
\begin{figure}
    \centering
    \includegraphics[width=\columnwidth]{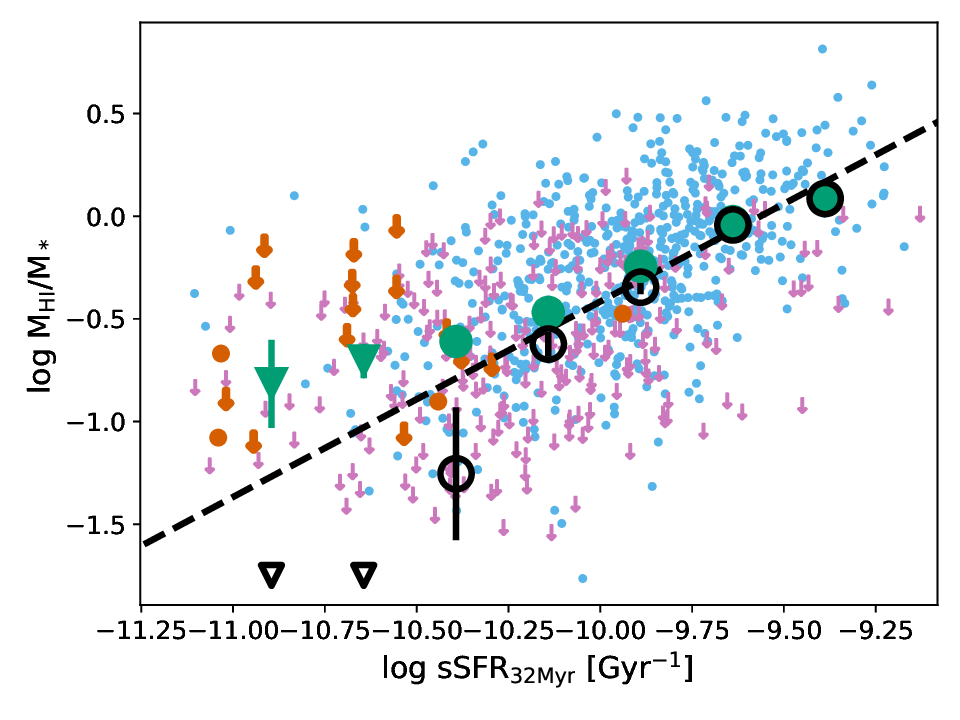}
    \caption{{\gs} vs. sSFR measured over the last 32 Myr. Symbols are the same as in Figure~\ref{fig:ew_gs}. This correlation is notably stronger than the relationship between {\gs} and EW(H$\alpha$) which traces SFR over $\sim$5 Myr timescales, supporting the idea that {\gs} is a strong influence on long-term averaged star formation.}
    \label{fig:ssfr32}
\end{figure}

We do not actually need to average over $\sim$Gyr timescales to see a tightening in the relationship between {\gs} and sSFR. Figure~\ref{fig:ssfr32} plots {\gs} vs. sSFR$_{\rm 32Myr}$, where sSFR$_{\rm 32Myr}$ is the sSFR over the last 32 Myr derived from \texttt{Pipe3d} stellar population synthesis modeling of the MaNGA datacubes. SFR$_{\rm 32Myr}$ is calculated by estimating the SFR history from fitting a composite stellar population to each spaxel spectrum,  taking into account the contribution of dust extinction. The obtained SFR history is then averaged within a period of 32 Myr for each spaxel, then coadded across the full field of view in order to obtain SFR$_{\rm 32Myr}$. Although sSFR$_{\rm 32Myr}$ does not trace star formation over the longer timescales often probed by color, it still averages over timescales several times that of the H$\alpha$ emission.  The resulting correlation is notably stronger ($\tau=0.35$) than the {\gs}-EW(H$\alpha$) relation. However, we must also consider the possibility that at least some of the scatter in the {\gs}-EW(H$\alpha$) relation is due to the aforementioned potential systematic uncertainties in the H$\alpha-$SFR calibration. Discrepancies between the assumed and true conditions of star-forming regions across a wide range of galaxies may contribute to a more scattered relation of {\gs} with H$\alpha$ emission, while we recover cleaner relation with sSFR$_{\rm 32Myr}$ where the SFR is based on stellar population modeling.   

\subsubsection{Why does $\text{M}_\text{HI}/\text{M}_*$ more strongly correlate with oxygen emission?}
\label{sec:gs_oxygen_why}

{\gs} is most strongly correlated with EW(O), and the correlation strengths are approximately equal regardless of which available oxygen line is considered. A correlation between \ion{H}{i} and [\ion{O}{i}] is expected to some extent, as oxygen has an ionization energy very similar to hydrogen (both approximately 13.6 eV), so they are expected to co-exist within the ISM. Indeed [\ion{O}{i}] is typically observed in partially neutral regions of the ISM like the outer edges of nebula \citep{Reynolds98,Kewley19}. However, the actual emissivity of the [\ion{O}{i}] line is closely tied to the excitation mechanism, not just its presence in neutral gas. Furthermore, the [\ion{O}{iii}] and [\ion{O}{ii}] lines are found throughout the inner regions of star-forming nebulae \citep{Kewley19} where hydrogen is mostly ionized, yet they show similarly strong correlations with {\it neutral} hydrogen.

The link between {\gs} and EW(O) can more simply be understood as a result of the relation between  {\gs} and $12+\log{O/H}$. Simply put, at high enough metallicities, oxygen line emission will {\it inversely} correlate with oxygen abundance as a result of the higher metallicity increasing the cooling efficiency of the gas and lowering its overall temperature (this trend reverses at low metallicity, but we are well above that regime; \citealt{McGaugh91}). EW(O) depends on the total line flux and the stellar continuum flux density, and as {\gs} increases, two changes occur. The metallicity decreases, thereby increasing the relative oxygen emission from gas in the appropriate conditions, and the stellar mass/surface density will on-average decrease\citep{Catinella13}, thus lowering the stellar continuum flux density and further boosting the EW. If we assume star forming regions are the primary source of optical line emission, the oxygen line luminosity can be expressed as
\begin{equation}
    \label{eq:ewo1}
    L_\ion{O}{}\propto\left(\frac{O}{H}\right)^n M_g\left(\frac{M_{\rm SF}}{M_g}\right)
\end{equation}
where $\frac{O}{H}$ is the oxygen abundance relative to hydrogen, $n$ is some value less than zero to reflect the anti-correlation between oxygen abundance and relative line strength, $M_{\rm SF}$ is the gas mass in star forming regions, and $M_g$ is the total gas mass.  Since $EW$ is the emission line flux normalized by the stellar continuum flux density, we can rewrite this as
\begin{equation}
\label{eq:ewo2}
    EW(O) \propto \left(\frac{O}{H}\right)^n\left(\frac{M_{\rm SF}}{M_*}\right) \propto \left(\frac{O}{H}\right)^n sSFR
\end{equation}
where we have assumed $M_{\rm SF}$ is proportional to SFR. The assumption that oxygen line emission is only coming from star-forming regions may be an over-simplification, as will be discussed in the following section.

\subsubsection{The fraction of diffuse gas}
\label{sec:role_of_dig}
Optical line emission will not only arise from star forming regions, but also from DIG (also known as the warm ionized medium, or WIM). The contribution of DIG to the overall emission line luminosity of a galaxy can be substantial.  For instance, \citet{Oey07} find on average $\sim$60\% of H$\alpha$ emission arises from DIG, with the fraction being highest for low surface brightness galaxies (they do not find any significant trend with \ion{H}{i} fraction, but also do not incorporate \ion{H}{i} non-detections into their analysis). Metal line emission has also been observed from DIG, and in some cases is enhanced relative to Balmer hydrogen emission (e.g., [\ion{O}{ii}]/H$\beta$, [\ion{O}{i}]/H$\alpha$, [\ion{N}{ii}]/H$\alpha$, [\ion{S}{ii}]/H$\alpha$; \citealt{Reynolds85a, Reynolds85b, Reynolds98, Haffner99, Hoopes03, Madsen06,Voges06,Zhang17}), which may imply an even larger fraction of the integrated metal line emission arises from the DIG.  

To account for DIG, Eq~\ref{eq:ewo2} can be modified to read:
\begin{equation}
    EW(O) = \left(\frac{O}{H}\right)^n\left[A\left(\frac{M_{\rm SF}}{M_*}\right) + B\left(\frac{M_{\rm DIG}}{M_*}\right)\right]
\end{equation}
where $A$ and $B$ are unknown constants, and M$_{\rm DIG}$ is the amount of emitting gas associated with DIG. We make no attempt to constrain the unknowns in this equation here, and they probably vary with galaxy properties (mass, surface brightness, etc.), but it highlights our key takeaways: (1) the oxygen emission is correlated with gas-phase metallicity (2) both star-forming clouds and DIG contribute to the overall emission. The relative contributions of HII regions and DIG to the {\gs}-EW(O) relation may not be fixed, and likely vary smoothly on average along the trend for it to remain well-behaved (such smooth variations should just be absorbed into the linear fit coefficients). Evidence of a growing contribution from DIG as {\hi} fraction increases is seen in the correlation between {\gs} and [\ion{O}{i}]/H$\alpha$ (and perhaps to a weaker extent, the anti-correlation with $\mu_{H\alpha}$). Additionally, in Section~\ref{sec:resid}, we found that the scatter in the {\gs} vs. EW(O) relations correlates with [\ion{O}{i}]/H$\alpha$ and $\mu_{H\alpha}$, which suggests the scatter may be driven by variations in the DIG fraction.  However, we caution that similar results could be explained by increased shock heating in more gas-rich galaxies. DIG line ratios may themselves be driven by shocks, but this is a subject of debate \citep{Haffner09}.

 If our observations can be explained by more DIG in gas-rich galaxies, one might also wonder if this implies the presence of more diffuse {\hi} as well. Analyses of the neutral content of the DIG implies that it is indeed mostly ionized, with neutral fractions $<10\%$ \citep{Reynolds98, Hausen02}.  However, kinematic studies find that DIG is almost always found to coincide in both space and velocity with \ion{H}{i}, specifically the warm neutral medium (WNM; the warmer, diffuse, more extended neutral hydrogen) as opposed to the cold neutral medium (CNM; the cooler, denser hydrogen clouds) \citep{Spitzer93, Reynolds95, Hartmann97,Howk03,Haffner09}. {\it Therefore, it is possible that galaxies with more DIG may also have a larger fraction of their HI residing in the WNM.} A further implication of this relationship would be that the most gas-rich galaxies (which show evidence of a higher DIG fraction) would host a larger WNM fraction. Similarly, the scatter in the {\gs}-EW(O) relation, which we attributed to variations in the fraction of DIG emission, may also reflect variations in the amount of {\hi} in the CNM vs. WNM phases.  
 
\begin{figure*}
    \centering
    \includegraphics[width=\columnwidth]{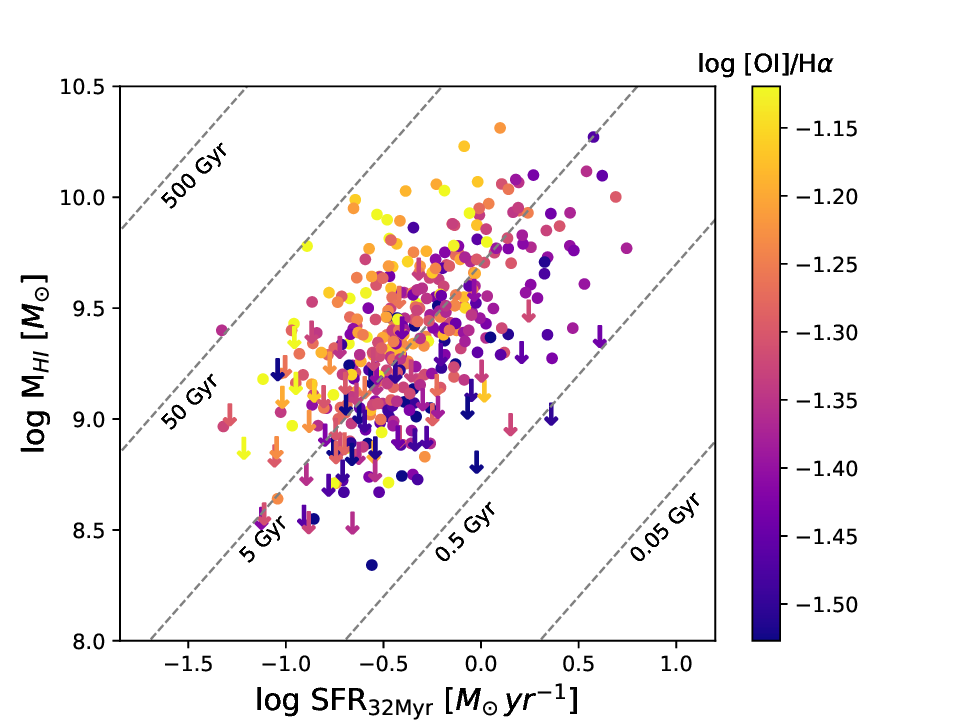}
    \includegraphics[width=\columnwidth]{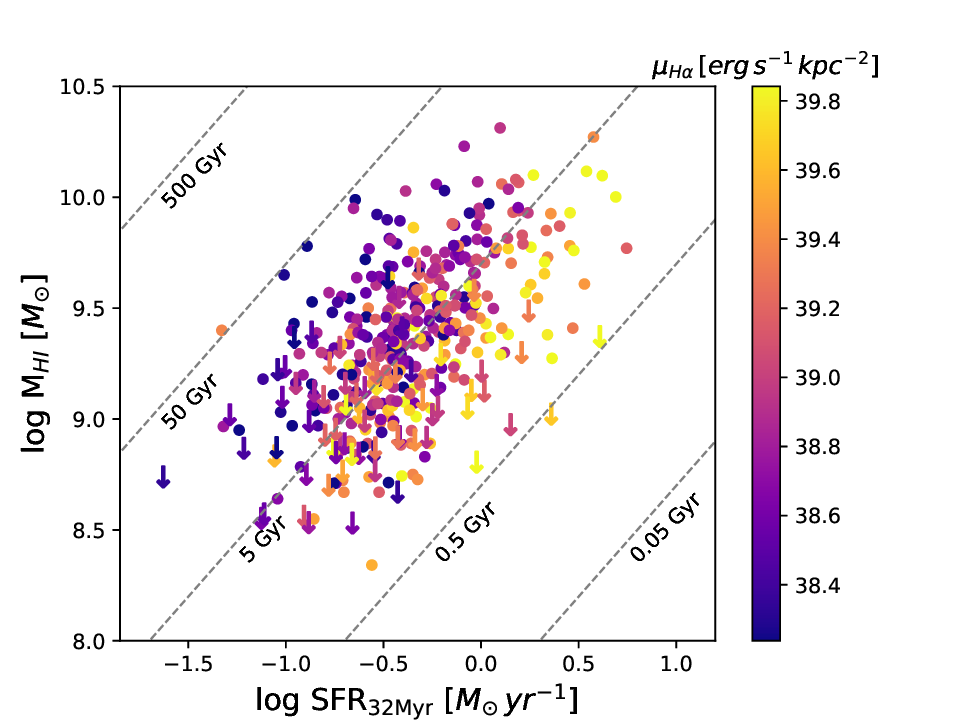}
    \caption{$M_{\hi}$ versus SFR for galaxies with $12+\log{O/H}<9.05$. Longer depletion times coincide with higher {\oiha} and depressed $\mu_{H\alpha}$, consistent with a larger fraction of DIG/WNM and/or gas heated by shocks.}
    \label{fig:tdep_oiha}
\end{figure*}

 Whether an {\hi} reservoir has a large diffuse component or is subject to more shock heating, such a change in the ISM may impact global star formation efficiencies. The {\hi} will not be able to condense into molecular clouds nearly as easily as it would otherwise, increasing the overall {\hi} depletion time, $t_{\rm dep}$, defined as $M_{\hi}/SFR$. We explore this possibility by plotting $M_{\hi}$ vs. SFR$_{\rm 32Myr}$ in Figure~\ref{fig:tdep_oiha} and comparing to lines of constant $t_{\rm dep}$. We specifically use SFR$_{\rm 32Myr}$ in order to smooth over stochastic variations in the SFR and look at the long-term averaged behavior. Points in Fig~\ref{fig:tdep_oiha} are color-coded by {\oiha} (left) and $\mu_{H\alpha}$ (right), and we restrict our analysis to the subset of galaxies with $12+\log{O/H}<9.05$ where there is no metallicity dependence on {\oiha}. Longer {\hi} depletion times at fixed SFR coincide with enhanced  {\oiha} and depressed $\mu_{\rm H\alpha}$, supporting the idea that longer {\hi} depletion times are caused by a larger fraction of {\hi} residing in the WNM and/or heated by shocks.

\section{Conclusions}
\label{sec:conclusions}
Using the second data release of the HI-MaNGA survey, an {\hi} follow-up program for the SDSS-IV MaNGA survey, we conducted an analysis of the scaling relations between galaxy {\hi}-to-stellar mass ratio and average ISM properties derived from optical spectroscopy for star forming galaxies.  Our key results are as follows:
\begin{itemize}

    \item We recover the strong anti-correlation between {\gs} and gas-phase metallicity seen in previous studies.  We also find a mild correlation between {\gs} and \oiha, which suggests the presence of a harder ionizing field, more shock excitation, and/or a larger fraction of diffuse ionized gas (DIG) in gas-rich galaxies.  A larger DIG fraction may in turn imply a larger fraction of {\hi} residing in the more diffuse warm neutral medium.

    \item {\gs} is weakly correlated with EW($H\alpha$) and other optical hydrogen lines. This weak connection implies the global {\hi} reservoir does not have a strong impact on the star formation on short-timescales. However, the stronger link between {\hi} and SFR when measured over longer timescales implies the {\hi} reservoir still plays an important role in sustaining the average SFRs of galaxies. 
    
    \item Of all optical emission lines, {\gs} correlates most strongly with Oxygen lines ([\ion{O}{i}], [\ion{O}{ii}], and [\ion{O}{iii}]). This result is likely driven by the existing anti-correlation between gas fraction and metallicity. 
    
    \item The residuals in the {\gs}-EW(O) relations are most strongly correlated with {\oiha} and mean H$\alpha$ surface brightness, $H\alpha$. This result suggests the scatter is driven by variations in the amount of gas associated with the warm neutral medium/DIG. 
    
    \item Galaxies with longer than average depletion times also have elevated {\oiha} and depressed $\mu_{H\alpha}$, consistent with long depletion times occurring when a significant fraction of {\hi} is in a diffuse phase and/or subject to more shock heating, making it more difficult to condense into molecular clouds.
    
    \item We make a first attempt to calibrate multi-parameter {\gs} scaling relations using optical spectroscopic information, but find that the asymmetric scatter makes them less preferable than existing scaling relations using broadband photometric properties. However, the relations in this work appear to strengthen, with notably less asymmetric scatter, if emission line properties are calculated using larger apertures.  More data with larger area coverage -- or careful aperture corrections applied to our current MaNGA sample -- may prove valuable when revisiting these scaling relations.
\end{itemize}

Our results highlight how ISM properties vary from typically low-mass gas-rich galaxies, to typically more massive gas-poor galaxies. In many ways, gas-rich dwarf galaxies represent aspects of the progenitors of the Milky Way and other more massive galaxies.  Understanding the properties of their ISM, and how they might impact star formation, is a key ingredient to our general understanding of galaxy formation.  Analyses of the relative fractions of diffuse and dense {\hi} may be particularly useful for simulations which aim to properly recreate the true breakdown of the ISM into its different phases.

In a practical sense, this work also demonstrates ways to conduct statistical analysis in the presence of {\hi} upper limits.  Both the Akritas-Theil-Sen estimator and substituting upper limits with photometric gas fractions are useful approaches and give generally consistent results.  A third approach, which we have not demonstrated here, is stacking of {\hi} data, with the caveat that stacking removes information about the distribution of data around mean trends. Regardless, {\hi} non-detections contain valuable information, it is crucial to incorporate them into any analysis of gas content in order to avoid biased results.

\section*{Acknowledgements}

We are very grateful to our referee for their careful review which improved our work. We also thank Eric Feigelson with for his very helpful advice on survival analysis techniques, and Christy Tremonti for her help interpreting our results. Rogerio Riffel thanks CNPq, CAPES and FAPERGS. VAR acknowledges financial support from CONACyT grant 285721. ZZ acknowledges support by NSFC grant U1931110. 

Funding for the Sloan Digital Sky 
Survey IV has been provided by the 
Alfred P. Sloan Foundation, the U.S. 
Department of Energy Office of 
Science, and the Participating 
Institutions. 

SDSS-IV acknowledges support and 
resources from the Center for High 
Performance Computing  at the 
University of Utah. The SDSS 
website is www.sdss.org.

SDSS-IV is managed by the 
Astrophysical Research Consortium 
for the Participating Institutions 
of the SDSS Collaboration including 
the Brazilian Participation Group, 
the Carnegie Institution for Science, 
Carnegie Mellon University, Center for 
Astrophysics | Harvard \& 
Smithsonian, the Chilean Participation 
Group, the French Participation Group, 
Instituto de Astrof\'isica de 
Canarias, The Johns Hopkins 
University, Kavli Institute for the 
Physics and Mathematics of the 
Universe (IPMU) / University of 
Tokyo, the Korean Participation Group, 
Lawrence Berkeley National Laboratory, 
Leibniz Institut f\"ur Astrophysik 
Potsdam (AIP),  Max-Planck-Institut 
f\"ur Astronomie (MPIA Heidelberg), 
Max-Planck-Institut f\"ur 
Astrophysik (MPA Garching), 
Max-Planck-Institut f\"ur 
Extraterrestrische Physik (MPE), 
National Astronomical Observatories of 
China, New Mexico State University, 
New York University, University of 
Notre Dame, Observat\'ario 
Nacional / MCTI, The Ohio State 
University, Pennsylvania State 
University, Shanghai 
Astronomical Observatory, United 
Kingdom Participation Group, 
Universidad Nacional Aut\'onoma 
de M\'exico, University of Arizona, 
University of Colorado Boulder, 
University of Oxford, University of 
Portsmouth, University of Utah, 
University of Virginia, University 
of Washington, University of 
Wisconsin, Vanderbilt University, 
and Yale University.

The Green Bank Observatory is a facility of the National Science Foundation operated under cooperative agreement by Associated Universities, Inc.

\section*{Data Availability}
The HI-MaNGA catalog used in this study can be found at  \url{https://greenbankobservatory.org/science/gbt-surveys/hi-manga/}. The MaNGA MPL-9 data products can be generated by the public using the raw data (available at \url{https://www.sdss.org/dr16/manga/manga-data/data-access/)} with DRP v2.7.1, DAP v2.4.3, and Pipe3D v2.7.1.







\appendix

\section{Performance of the Akritas-Theil-Sen Estimator}
\label{sec:app_ast}
\begin{figure}
    \centering
    \includegraphics[width=\columnwidth]{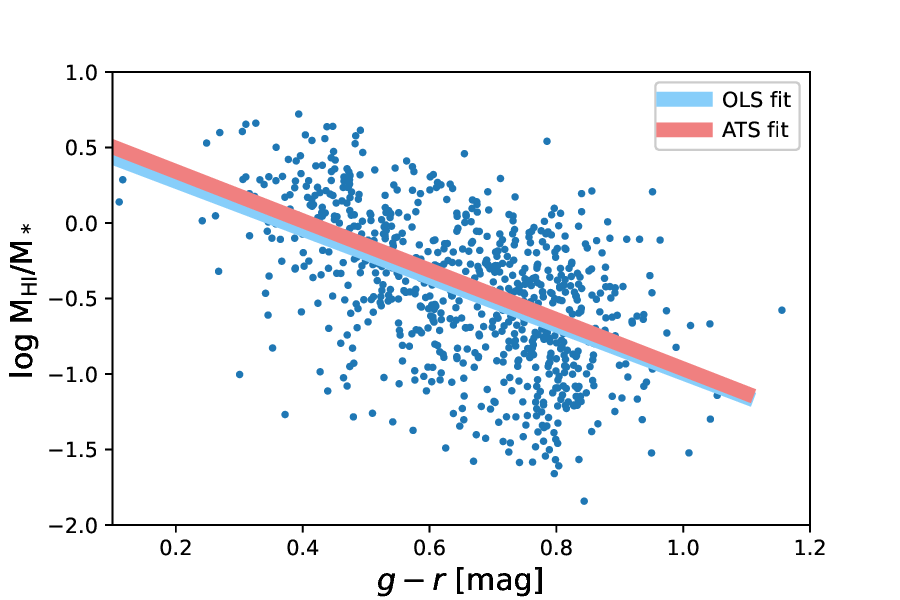}
    \caption{The {\gs} vs. $g-r$ relation using only \ion{H}{i} detections from the xGASS survey \citep{Catinella18}.  The red and blue lines represent the OLS and ATS fits to the data.}
    \label{fig:gs_color_xgass_nolim}
\end{figure}

 To test the performance of the ATS estimator, we examine its consistency and accuracy when applied to a mock data set where the fraction of censored data is progressively increased.  For this analysis we use data from the extended GALEX Arecibo SDSS xGASS Survey \citep[xGASS;][]{Catinella18}, chosen because it has deep \ion{H}{i} observations and a roughly uniform stellar mass distribution similar to our MaNGA data set. We start with only the xGASS detections, so that we have a sample where true {\hi} masses are known for all galaxies.  Fig.~\ref{fig:gs_color_xgass_nolim} shows the {\gs} vs. $g-r$ relation for this sample, as well as the ordinary unweighted least-squares (OLS) fit and the ATS fit.  These two fitting algorithms are in good agreement within their errors. 

The {\gs} vs. $g-r$ relation is a ideal test of the ATS estimator for this work because when we simulate different observing depths and introduce mock upper limits, the distribution of upper limits is similar to the distribution seen in the relations presented in Section~\ref{sec:results}. Specifically, the range of {\gs} for detections and non-detections overlaps significantly at fixed $x$-axis value. To create mock non-detections in this data set in a manner consistent with the non-detections in the HI-MaNGA survey, we assign each xGASS galaxy a new distance based on its $i$-band absolute magnitude, $M_i$, mimicking the relationship between $M_i$ and redshift which is explicitly built into the MaNGA Primary sample design \citep{Wake17}. For a given $M_i$, a galaxy in MaNGA will only fall within a set range of redshifts, and we assign each galaxy a random value within that range. To ensure consistency with MaNGA data, we use the ellipical Petrosian magnitudes from the NSA catalog.  Each galaxy's observed \ion{H}{i} flux is recalculated using its known \ion{H}{i} mass and this new distance.  We then assume all galaxies are observed down to a fixed $rms$ noise level, which we use to calculate the integrated flux $S/N$ for each galaxy.  Anything with a mock $S/N<3$ is assumed to be a non-detection and given a 3$\sigma$ upper limit assuming a linewidth of 200 ${\rm km\,s^{-1}}$. We start with an rms of 1.5 mJy (the nominal HI-MaNGA survey depth) and progressively increase it up to 10 mJy, each time running an OLS fit on the mock detections and an ATS fit on all the data, including the upper limits.

\begin{figure*}
    \centering
    \includegraphics[width=2\columnwidth]{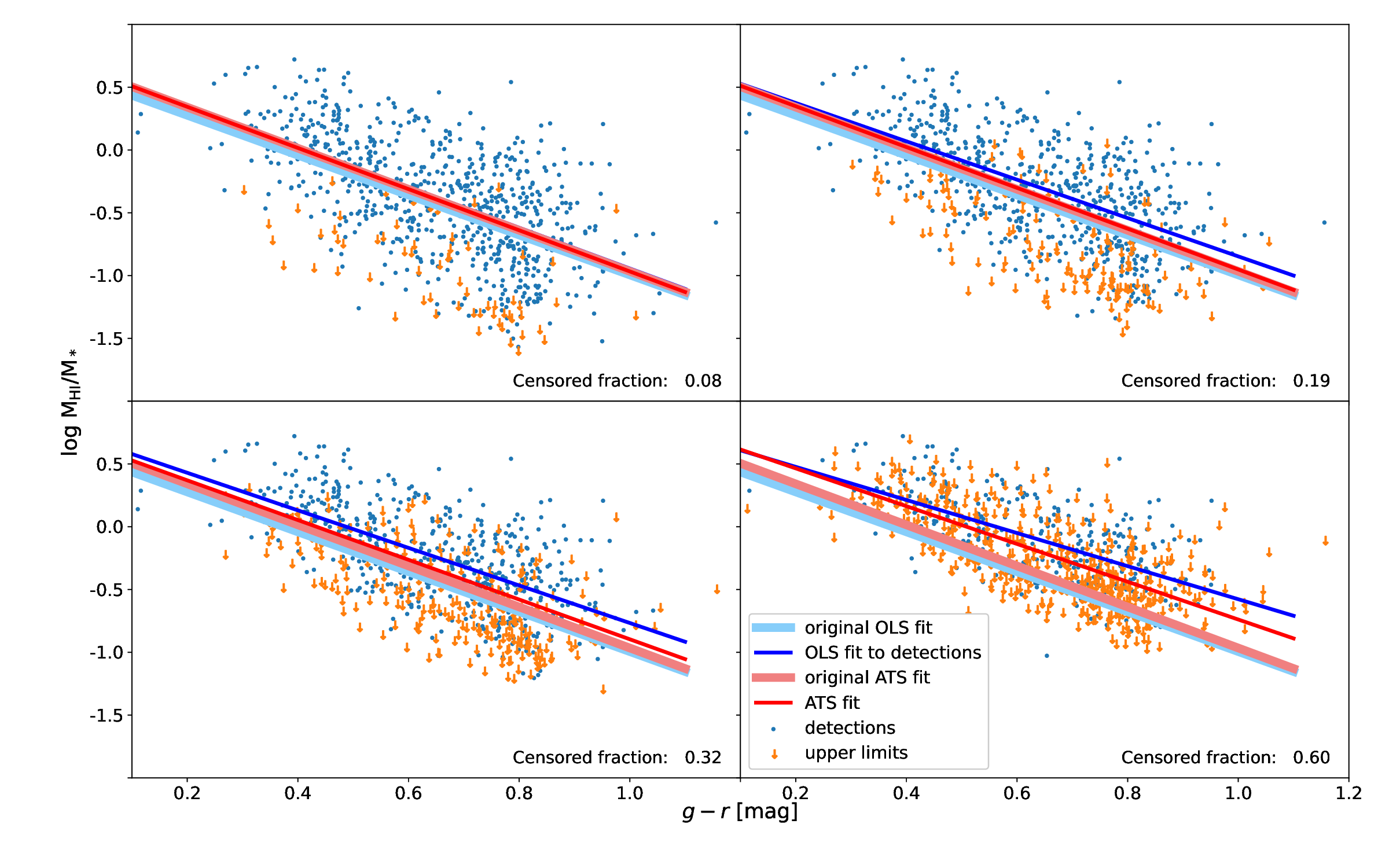}
    \caption{{\gs} vs. $g-r$ color relation for xGASS data where galaxies are redistributed in redshift space to match the MaNGA survey design, then mock observed to progressively lower depths. The thick lines represent the original fits to the uncensored data from Figure~\ref{fig:gs_color_xgass_nolim}. The Akritas-Theil-Sen line remains extremely stable until the censoring fraction exceeds $\sim$50\%.}
    \label{fig:gs_color_xgass_lim}
\end{figure*}

Figure~\ref{fig:gs_color_xgass_lim} shows the results of this analysis.  Unsurprisingly, the OLS fit on just the detections becomes increasingly biased as the mock survey depth decreases.  Meanwhile, the ATS estimator fit changes very little, until the fraction of censored data exceeds $\sim$50\%, at which point the ATS fit is clearly biased too high and shallow relative to the true relation, although the ATS is less biased than an OLS fit to just the detections. Notably the ATS estimator and the true fit are in good agreement at a censoring fraction of $\sim$30\%, the same censoring fraction as our analysis sample. This analysis illustrates how the ATS estimator provides both a consistent and accurate linear fit through the data even in the presence of significant censoring.  

\section{Photometric Gas Fraction Calibration}
\label{app:photogs}

To determine $P({\rm M_{\hi}/M_*}|{\rm color})$, the 2D probability distribution of galaxies as a function of {\gs} and modified color, we follow the methodology of \citet{Eckert15} using the RESOLVE survey {\hi} catalog \citep{Stark16}. RESOLVE is the ideal data set to determine this probability distribution as it is a highly complete, closed volume survey with {\hi} data that is uniformly deep (as a fraction of stellar mass). We select only data from the  RESOLVE-A volume above the baryonic mass completeness limit, enforce all detections to have $S/N>5$, and reject any confused targets whose systematic error due to confusion is $>25\%$ of the integrated flux. The RESOLVE catalog is crossmatched with the NSA using a search radius of $5\arcsec$ in order for us to use photometry and stellar masses consistent with our MaNGA sample.

We first determine which color provides the tightest correlation with {\gs} when using NSA data, which we find to be $u-i$.  We next determine the optimum third parameter, testing both axial ratio ($b/a$) and $r$-band surface brightness calculated within $R_e$ ($\mu_r$).  The fits are limited to the {\hi} detections between $0.5 < u-i < 3$ in order to avoid very red colors where the data are dominated by upper limits and very blue colors where the scatter appears asymmetric. While \citet{Eckert15} find $b/a$ to be the optimal third parameter, we find it has very little correlation with the residuals in the \gs-$(u-i)$ correlation. A more significant correlation is found with $\mu_r$.  Based on the fit between \gs, $u-i$, and $\mu_r$, our final modified color is defined as $0.878(u-i)-0.116\mu_r+4$.  The additive factor of 4 is simply to ensure the final modified color is $>0$, which is needed for the full fit to $P({\rm M_{\hi}/M_*}|{\rm color})$.
 
$P({\rm M_{\hi}/M_*}|{\rm color})$ is fit using Eqs. 1--4 from \citet{Eckert15}. To briefly summarize, the model assumes there are two distinct populations: (1) {\hi}-rich galaxies (typically detections) which follow a linear relation with modified color and have Gaussian scatter which broadens at redder colors, and (2) gas-poor (typically non-detections) which are assumed to all fall around {\gs}$\sim0.05$. We use {\gs} and color bin sizes of 0.2 and weight the data by $1/N$, where $N$ is the number of galaxies per bin.  The final fit parameters are given in Table~\ref{tab:pgs_fit}.  

For each non-detection in our MaNGA sample, we determine its {\gs} probability distribution based on its modified color and the model fit parameters given in Table~\ref{tab:pgs_fit}. We set all probabilities to zero above the measured upper limit.  The new estimate of {\gs} for this galaxy is determined by randomly drawing from its {\gs} probability distribution.  

\begin{table}
    \centering
    \caption{Best fit values to the $P(M_{\ion{H}{i}}|{\rm color})$ model}
    \begin{tabular}{c|c}
    \hline
       $A_0$  &  26.38\\
       $A_1$  &  0.93\\
       $A_2$  &  0.15\\
       $A_3$  &  -1.10\\
       $A_4$  &  2.69\\
       $A_5$  &  0.12\\
       $A_6$  &  46.52\\
       $A_7$  &  3.78\\
       $A_8$  &  0.37\\
       \hline
    \end{tabular}
    \label{tab:pgs_fit}
\end{table}


\bsp	
\label{lastpage}
\end{document}